\documentclass[aps, prx,reprint,twocolumn,superscriptaddress,showpacs,floatfix]{revtex4-2}

\usepackage{amsmath, amssymb, amsfonts, amsthm, bm, bbm, dsfont, upgreek, mathtools} 
\usepackage{braket} 
\usepackage[vcentermath]{youngtab} 
\usepackage[nottoc,notlot,notlof,notbib]{tocbibind} 
\usepackage{graphicx} 
\usepackage{multirow, makecell} 
\usepackage{pgf, epstopdf} 
\usepackage{float}

\usepackage{times} 
\usepackage{lipsum} 
\usepackage{enumitem} 
\usepackage[normalem]{ulem} 
\usepackage{textcomp} 
\usepackage{ragged2e} 
\usepackage{hyperref} 
\usepackage{nameref} 
\hypersetup{
    colorlinks = true,
    linkcolor = [rgb]{0.70,0.13,0.13},
    citecolor = [rgb]{0.13,0.55,0.13},
    urlcolor = [rgb]{0.25, 0.41, 0.88}
}

\newtheoremstyle{example}
  {10pt} {10pt} {} {} {\bfseries} {.} {5pt plus 1pt minus 1pt} {}
\newtheoremstyle{preliminary}
  {10pt} {10pt} {} {} {\bfseries} {.} {5pt plus 1pt minus 1pt} {}

\theoremstyle{plain}

\theoremstyle{example}

\definecolor{lightRed}{RGB}{255, 182, 193}
\newcommand{\ii}{\mathrm{i}}

\newcommand{\Tr}{\mathop{\operatorname{Tr}}}

\definecolor{purple}{rgb}{.6,.1,.6}
\definecolor{darkgreen}{rgb}{.1,.6,.1}

\begin{document}

\title{Quantum Strong-to-Weak Spontaneous Symmetry Breaking in Decohered One Dimensional Critical States}

\author{Yuxuan Guo}
\altaffiliation{The first two authors contributed equally.}
\email{yuxguo2024@g.ecc.u-tokyo.ac.jp}
\affiliation{Department of Physics, University of Tokyo, 7-3-1 Hongo, Bunkyo-ku, Tokyo 113-0033, Japan}

\author{Sheng Yang}
\altaffiliation{The first two authors contributed equally.}

\affiliation{Institute for Advanced Study in Physics and School of Physics, Zhejiang University, Hangzhou 310058, China}

\author{Xue-Jia Yu}
\email{xuejiayu@fzu.edu.cn}
\affiliation{Department of Physics, Fuzhou University, Fuzhou 350116, Fujian, China}
\affiliation{Fujian Key Laboratory of Quantum Information and Quantum Optics,
College of Physics and Information Engineering,
Fuzhou University, Fuzhou, Fujian 350108, China}

\begin{abstract}
Symmetry breaking has been a central theme in classifying quantum phases and phase transitions. Recently, this concept has been extended to the mixed states of open systems, attracting considerable attention due to the emergence of novel physics beyond closed systems. In this work, we reveal a new type of phase transition in mixed states, termed \emph{quantum} strong-to-weak spontaneous symmetry breaking (SWSSB).  Using a combination of field theory calculations and large-scale matrix product state simulations, we map out the global phase diagram of the XXZ critical spin chain under local strong symmetry preserving decoherence, which features an SWSSB phase and a trivial Luttinger liquid phase, separated by a straight critical line that belongs to the boundary Berezinskii-Kosterlitz-Thouless universality class with a varying effective central charge.  Importantly, we analyze this transition from two complementary perspectives: on one hand, through the behavior of order parameters that characterize the symmetry breaking; on the other hand, from a quantum information viewpoint by studying entropic quantities and the concept of quantum recoverability. This dual approach allows us to provide a more comprehensive understanding of the phase structure and the nature of the transition. Remarkably, the SWSSB transition in our case is \emph{purely quantum} in the sense that it can only be driven by tuning the Hamiltonian parameter even under arbitrary decoherence strength, fundamentally distinguishing it from the decoherence-driven SWSSB transitions extensively discussed in previous literature. Importantly, our unified theoretical framework is applicable to a broad class of one-dimensional quantum systems, including spin chains and fermionic systems, whose low-energy physics can be described by Luttinger liquid theory, under arbitrary symmetry-preserving decoherence channels. Finally, we also discuss the experimental relevance of our theory on quantum simulator platforms.

\end{abstract}

\maketitle
\date{\today}

\section{Introduction}
The notion of symmetry forms a cornerstone in understanding quantum phases and phase transitions in modern physics~\cite{landau1980statistical}. Typically, different phases are distinguished by their symmetry patterns, and transitions between them can be described by spontaneous symmetry breaking both within and beyond the Landau paradigm~\cite{RevModPhys.89.041004,sachdev1999quantum,mcgreevy2023generalized,zheng2024experimentaldemonstrationspontaneoussymmetry}. Furthermore, symmetry and its breaking play a central role in cosmology and particle physics, influencing phenomena ranging from the evolution of the universe~\cite{KIBBLE1980183,volovik2003universe} to the origin of the mass of fundamental particles~\cite{anderson1958,higgs1964a,higgs1964b,nambu1961,Nambu2009}.  

Recent progress~\cite{Lu2020PRL,lee2022decodingmeasurementpreparedquantumphases,Lu2022PRXQ,Ma2023prx,ma2023topological, Lu2023PRXQ,PhysRevX.13.031016, PhysRevX.8.011035, PRXQuantum.5.030304, PRXQuantum.6.010315,PhysRevLett.134.096503,stephen2024manybodyquantumcatalyststransforming,sun2025holographicviewmixedstatesymmetryprotected,zhang2024fluctuationdissipationtheoreminformationgeometry,guo2024locallypurifieddensityoperators,xue2024tensornetworkformulationsymmetry,Wang2025prxq,wang2024anomalyopenquantumsystems,dai2023steadystatetopologicalorder,wang2023topologicallyorderedsteadystates,lessa2024mixedstatequantumanomalymultipartite,wang2024analogtopologicalentanglemententropy,You2024prb,chen2024unconventionaltopologicalmixedstatetransition,Chen2024PRL,yu2025gaplesssymmetryprotectedtopologicalstates,zhang2025universalpropertiescriticalmixed,zhou2025finitetemperaturequantumtopologicalorder,myersonjain2024pristinepseudogappedboundariesdeconfined,Su2024prb,Su2024PRL,ma2024symmetryprotectedtopologicalphases,ma2023exploringcriticalsystemsmeasurements,su2025spinliquidsuperconductivityemerging,lessa2025higherformanomalylongrangeentanglement,sarma2025effectiveconformalfieldtheory,Feng2025Hardness} has revealed that the classification of quantum phases extends to mixed states in open systems, attracting significant interest from both condensed matter communities and quantum information science \cite{sang2024stability,kwon2022reversing,PhysRevX.14.031044,Xu2025PRB}. In this context, understanding the symmetry properties of open quantum systems is crucial for uncovering their emergent behaviors and has long been elusive. The conventional understanding of symmetry in closed quantum systems requires fundamental revision when applied to open systems, where the interplay between the system and its environment leads to a richer classification stemming from the distinct nature of mixed states. It has been pointed out that there are two distinct notions of symmetry for open quantum systems~\cite{Buca:2012zz,PhysRevA.89.022118,PhysRevLett.125.240405}. For a density matrix $\hat{\rho}$ representing a pure state $\ket{\psi}$ that is invariant under a symmetry group $G$, the invariance of the density matrix under both left and right unitary operations $\hat{U}_g$ is expressed as $\hat{U}_g \hat{\rho} = (\hat{\rho} \hat{U}_g^\dagger )^\dagger= e^{\ii\phi}\hat{\rho}$, where $g\in G$ and $\hat{U}_g$ is a nontrivial representation of $G$. This is commonly referred to as \emph{strong symmetry}. In contrast, for mixed states, symmetry can persist only at the level of the ensemble average, even when strong symmetry is absent. In such cases, the symmetry is described by the condition $\hat{U}_g \hat{\rho} \hat{U}_g^\dagger = \hat{\rho}$, which is often termed \emph{weak symmetry}. In this context, symmetry can transition from being intact at the pure-state level to being preserved only at the ensemble average level. This phenomenon, referred to as \emph{strong-to-weak spontaneous symmetry breaking} (SWSSB)  
 \cite{lessa2024strong,sala2024spontaneous,bao2023mixed, PRXQuantum.5.020343,huang2024hydrodynamics,gu2024spontaneous,kuno2024strong,kuno2025strongtoweakspontaneoussymmetrybreaking,zhang2024strong,Orito:2024cel,Kuno:2025mcr,kuno2025intrinsicmixedstatetopological,PhysRevB.111.L060304,weinstein2024efficientdetectionstrongtoweakspontaneous,kim2024errorthresholdsykcodes,ando2024gaugetheorymixedstate,PhysRevLett.131.220403,PRXQuantum.4.030317,Shah2024Instability}
 , is an intrinsic feature of mixed states and is fundamentally distinct from the traditional notion of symmetry breaking, which transitions from preserving strong symmetry directly to the absence of symmetry. 

However, the majority of theoretical studies to date have induced
SWSSB by {increasing}
the decoherence.  Strictly local dephasing channels
\cite{Gong2024}, or more generally any dephasing that lacks a protected
non-Abelian structure, ultimately disentangle the system; one therefore
expects a decoherence–driven SWSSB transition, a scenario that has been
extensively analysed in the literature.  In practice, however, the strength of decoherence is notoriously difficult to tune and even more challenging to measure with precision. Moreover, such tuning and measurement often rely on prior assumptions about the noise model, which poses significant obstacles to accurately observing and characterizing SWSSB phase transition in open quantum systems experimentally. This limitation prompts a fundamental question:

  \emph{Can SWSSB occur in a genuinely quantum regime—namely, one in
  which the transition is driven by Hamiltonian parameters which is much easier to tune in real material or quantum platform while the
  environment is symmetric?}

Such a situation is the open–system analogue of a conventional quantum
phase transition: criticality is reached by varying couplings \emph{in
the Hamiltonian}, rather than by adjusting the system–bath coupling
(which plays a role analogous to temperature) and the phase transition should be insensitive to the details of system bath coupling.

In this work we answer the above question affirmatively.  We uncover a
\emph{quantum SWSSB transition} in a critical
one-dimensional model that, in the absence of noise, exhibits the
characteristic logarithmic scaling of entanglement.  Coupling the system
to local decoherence channels that \emph{exactly} preserve the global
symmetry reveals two distinct phases:

\begin{enumerate}[label=(\alph*)]
  \item For one region of Hamiltonian parameters, the strong symmetry of
        the underlying pure state is robust against \emph{arbitrarily
        large} symmetry-respecting noise.
  \item For another region, an \emph{infinitesimal} amount of the same
        noise is sufficient to destroy the symmetry.
\end{enumerate}
A continuous phase boundary separates the two regimes.
Remarkably, the associated critical point is intrinsic to the open quantum system and has \emph{no counterpart in the closed system} when the decoherence goes to zero; it cannot be interpreted as any form of phase transition of a pure state under decoherence. Moreover, this transition can be accessed with \emph{arbitrarily} symmetry-preserving decoherence.  In this respect the phenomenon serves
as the open–system counterpart of a conventional zero-temperature
quantum phase transition: criticality is attained by tuning Hamiltonian
parameters—thereby changing the ground-state structure—rather than by
varying the strength of the system–bath coupling. To the best of our
knowledge, this class of quantum phase transitions has not been reported
previously. Our analysis combines an effective-field-theory description with
large-scale numerical simulations, thereby elucidating the subtle
interplay between quantum criticality and decoherence and extending the
study of SWSSB far beyond stabilizer states with area-law entanglement.


\begin{figure}[tb]
    \centering
    \includegraphics[width=0.85\linewidth]{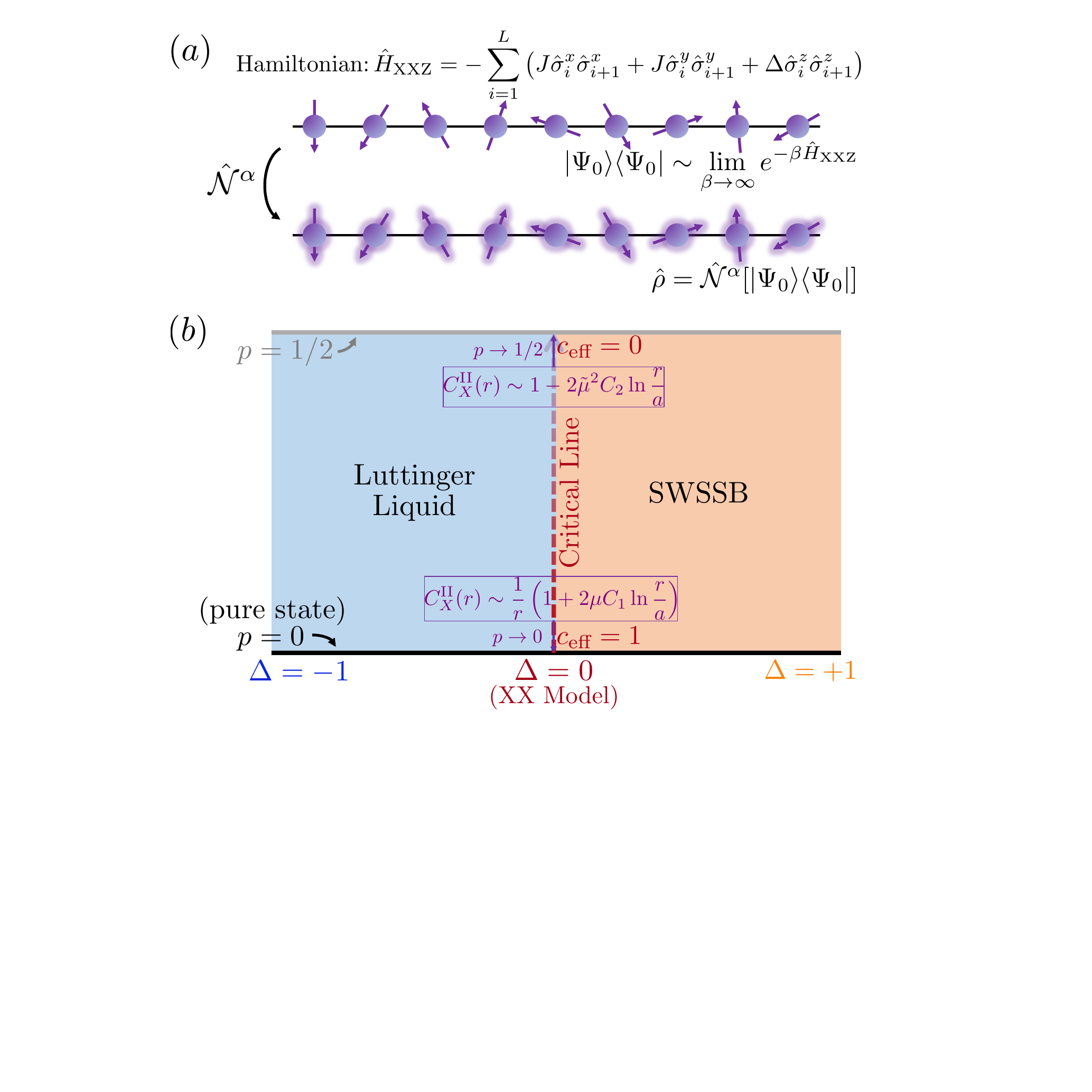}
    \caption{(a) Schematic diagram of our model describing the ground state of Hamiltonian~\eqref{eq:Hxxz} under decoherence. (b) Phase diagram of the XXZ model under XX decoherence. The system exhibits a boundary BKT phase transition at the critical point $\Delta_c = 0$. For $\Delta < 0$, the system is in a trivial Luttinger liquid phase with strong symmetry, whereas for $\Delta > 0$, it transitions into the SWSSB phase, where only weak symmetry is preserved. At the critical point $\Delta_c = 0$, the correlation length diverges as $\xi \sim (\Delta - \Delta_c)^{-\nu}$ with $1/\nu = 0$, consistent with the boundary BKT universality class. The red dashed critical line is characterized by an effective central charge $c_{\text{eff}}$ of the Choi state $|\rho\rangle\rangle$ and critical exponents of R\'enyi-2 correlation function, which decrease with increasing decoherence.}
    \label{fig:phase_diagram}
\end{figure}

The manuscript is structured as follows: Section~\ref{section:2} establishes the theoretical foundation, presenting the XXZ spin chain Hamiltonian and its low-energy effective description as a Luttinger liquid with tunable Luttinger parameter, while defining the order parameters essential for characterizing the emergent phases. In Section~\ref{section:3}, we develop a comprehensive field-theoretic framework by combining the Choi isomorphism with bosonization techniques to systematically analyze the SWSSB mechanism in Luttinger liquids subject to decoherence. Sections~\ref{section:4.1} and~\ref{section:4.2} present a comprehensive renormalization group analysis that systematically maps the quantum phase diagram of the XXZ model with XX decoherence, parametrized by anisotropy and decoherence strength. Our analysis identifies a critical line at $\Delta_c=0$ (equivalently, Luttinger parameter $K_c=1$) demarcating two distinct quantum phases, with the transition manifesting boundary BKT universality class . Through detailed entanglement measures, we provide robust validation of these phases and their distinctive properties. A remarkable feature of this transition is its purely quantum nature—it can be induced exclusively by tuning a Hamiltonian parameter even at arbitrary decoherence strength. Section~\ref{section:4.3} discuss the local recoverablity and universal recovery map of different phases. Section~\ref{section:4.4} explores the critical properties along the $\Delta_c=0$  line across the full spectrum of decoherence strengths, revealing non-trivial scaling behavior in both the effective central charge and critical exponents under marginal $XX$ decoherence. Importantly, the analytical framework developed in Section~\ref{section:4} extends naturally to other decoherence channels, predicting analogous boundary BKT-like phase transitions at different critical values of the Luttinger parameter. Section~\ref{section:5} offers a similar analysis of ZZ decoherence, where renormalization group calculations demonstrate that the system maintains its Luttinger liquid phase without SWSSB within the parameter range representable by the critical XXZ model—a finding corroborated by our numerical simulations. Section~\ref{section:gen} discusses the broader applicability of our methodology to generic decoherence channels, extending beyond the specific decoherence types examined in our detailed analyses of the XXZ model. Finally, in Section~\ref{section:6}, we summarize the key results of this work and explore potential experimental realizations of the SWSSB phenomenon.

\section{Model and method}
\label{section:2}
A particularly illuminating platform for studying SWSSB is the spin-$1/2$ quantum XXZ model, described by the Hamiltonian \cite{giamarchi2004quantum} (see Fig.~\ref{fig:phase_diagram} (a)):
\begin{gather}
    \hat{H}_{\text{XXZ}} = -\sum_{i=1}^L \left( J\hat{\sigma}^x_i \hat{\sigma}^x_{i+1} + J\hat{\sigma}^y_i \hat{\sigma}^y_{i+1} + \Delta \hat{\sigma}^z_i \hat{\sigma}^z_{i+1} \right) \, ,
    \label{eq:Hxxz}
\end{gather}
where $\hat{\sigma}^{\alpha}_i$ ($\alpha = x, y, z$) denotes the Pauli operators at site $i$, and $\Delta$ characterizes the anisotropy in the $z$-direction. For simplicity, we set $J=1$ as the energy unit throughout the work. This model is of particular interest for two reasons. First, it represents a minimal model for frustrated spin systems, extending beyond previous studies starting from \emph{area-law entangled} stabilizer states~\cite{lessa2024strong,guo2024strong,sala2024spontaneous,zhang2024strong}, where the underlying physics of these studies is often interpreted through mappings to corresponding classical lattice statistical mechanics models~\cite{lessa2024strong,sala2024spontaneous,PRXQuantum.5.020343,PhysRevB.111.115123,zhang2024strong,lee2025symmetry,PhysRevB.109.195420,liu2024noise,PhysRevB.110.085158,PhysRevA.111.032402}. Second, its low-energy physics is captured by the Tomonaga-Luttinger liquid theory with the action:
\begin{gather} 
\label{eq:LL_action}
\mathcal{S}_\text{XXZ} = \frac{1}{2\pi} \int dx \, d\tau \left[ i (\partial_\tau \phi)(\partial_x \theta) + \frac{1}{K} (\partial_x \phi)^2 + K (\partial_x \theta)^2 \right].
\end{gather}
Here, $\theta$ and $\phi$ are dual bosonic fields with periodicities $2\pi$ and $\pi$, respectively, satisfying the canonical commutation relation $[\partial_x \hat{\theta}(x), \hat{\phi}(x')] = i\pi\delta(x-x')$. The Luttinger parameter $K = \frac{\pi}{2\arccos{\Delta}}$ can be continuously tuned, and the action~\eqref{eq:LL_action} describes the low-energy behavior of a wide range of one-dimensional quantum systems.

Starting from the ground state $\ket{\Psi_0}$ of the XXZ chain, we model decoherence via a composition of local quantum channels acting on neighboring sites through two-qubit Pauli operators:
\begin{align}
\hat{\mathcal{N}}^\alpha_i[\ket{\Psi_0}\!\bra{\Psi_0}] &= 
(1-p)\ket{\Psi_0}\!\bra{\Psi_0} \nonumber \\
&\quad + p \hat{\sigma}_i^\alpha \hat{\sigma}_{i+1}^\alpha 
\ket{\Psi_0}\!\bra{\Psi_0} 
\hat{\sigma}_i^\alpha \hat{\sigma}_{i+1}^\alpha \, ,
\end{align}
where $\alpha \in \{x,z\}$. These channels represent minimal choices that preserve strong   $\mathbb{Z}_2^{x/y/z}$ symmetries, with $\mathbb{Z}_2^{x/y/z}$ representing $\pi$ rotations along the $x/y/z$ axis generated by $\hat{U}_{x/y/x} = \prod_{i=1}^L \hat{\sigma}_i^{x/y/z}$. Since YY decoherence $\hat{\sigma}_i^y \hat{\sigma}_{i+1}^y$ is equivalent to XX decoherence under a rotation in $x-y$ plane, we will only foucs on $XX$ decoherence in this article. Here, $0 \leq p < \frac{1}{2}$ characterizes the strength of decoherence. And the relation of those two-sites channel with general local strong symmetric channels will be discussed in section~\ref{section:gen}. By applying these local quantum channels together, we obtain the resulting mixed state:
\begin{gather}
\hat{\rho} = \hat{\mathcal{N}}^{\alpha}[\ket{\Psi_0}\!\bra{\Psi_0}] = \prod_{i=1}^L \hat{\mathcal{N}}^\alpha_i[\ket{\Psi_0}\!\bra{\Psi_0}] \, .
\end{gather}
Since quantum states transformed by finite-depth local quantum channels cannot always be reversed via other local quantum channels, even pure states may exhibit emergent quantum phases under such transformations—a phenomenon fundamentally distinct from the behavior of finite-depth local unitaries in closed systems. Although the quantum channel preserves strong symmetry (i.e. all Karus operators commute with symmetry operations, see Appendix.~\ref{appendix:A.1} for details), the decohered state $\hat{\rho}$ can exhibit spontaneous breaking of strong symmetry to weak symmetry. Traditional correlation functions, $\Tr[\hat{\rho} \hat{O}^\dagger_i \hat{O}_j]$, fail to capture this transition because the operators $\hat{O}$ (e.g., $\hat{\sigma}^x_i$ or $\hat{\sigma}^z_i$ in this work) are charged under weak symmetry and thus cannot detect SWSSB. To characterize different phases, we mainly employ the R\'enyi-2 correlator:
\begin{gather}
    C^{\text{II}}_{O}(j-i) = \frac{\Tr[\hat{\rho} \hat{O}^\dagger_i \hat{O}_j \hat{\rho} \hat{O}^\dagger_j \hat{O}_i]}{\Tr (\hat{\rho}^2)} \, .
\end{gather}
This correlator proves particularly effective in distinguishing between different phases in our system. While our analysis primarily focuses on the Rényi-2 correlation functions \cite{zhang2024strong,su2024emergent}, we acknowledge that alternative correlators can also effectively differentiate between the trivial phase and the SWSSB phase \cite{lessa2024strong,weinstein2024efficient,liu2024diagnosing}. Readers unfamiliar with these topics can find a comprehensive summary explaining the efficacy of Rényi-2 and other correlation functions for SWSSB detection in Appendix~\ref{appendix:A.2}.

\section{Effective field theory of the Choi state}
\label{section:3}
To establish a field-theoretical framework for mixed states, we use the Choi-Jamiołkowski isomorphism to rewrite the density matrix $\hat{\rho} = \sum_{i,j} \rho_{ij} \ket{i}\!\bra{j}$ as a double state $\left| \rho \right\rangle\rangle = \sum_{i,j} \rho_{ij} \ket{i}_{L} \ket{j}^*_R$, where $L,R$ label the two pieces of the Choi state $\left| \rho \right\rangle\rangle$. Within this representation, the R\'enyi-2 correlator takes the explicit form:
\begin{gather}
    C^{\text{II}}_{O}(i-j)=\frac{\langle\langle\rho|\hat{O}_{L;i}\hat{O}^*_{R;i}\hat{O}^\dagger_{L;j}\hat{O}^T_{R;j}|\rho\rangle\rangle}{\langle\langle\rho|\rho\rangle\rangle} \, .
\end{gather}
And it is useful to treat the quantum channel as an imaginary-time evolution in the doubled Hilbert space to facilitate field theory:
\begin{gather}
    \left| \rho \right\rangle\rangle = \cosh^{-L}(u) e^{u \sum_{i=1}^{L} \hat{\sigma}_{L,i}^\alpha\hat{\sigma}_{L,i+1}^\alpha \hat{\sigma}_{R,i}^\alpha\hat{\sigma}_{R,i+1}^\alpha} \ket{\Psi_0}_L \ket{\Psi_0}_R,
\end{gather}
where $u = \tanh^{-1}\left[p/(1-p)\right]$ parametrizes the decoherence strength. 

This formalism introduces a boundary term in the effective action:
$
    \mathcal{S}_d^\alpha = 2u \sum_{i=1}^L \int d\tau\ \delta(\tau)\ {\sigma}_{L;i}^\alpha {\sigma}_{L;i+1}^\alpha {\sigma}_{R;i}^\alpha {\sigma}_{R;i+1}^\alpha,
$
which encodes the competition between coherent dynamics and decoherence, and in fact $\delta(\tau)$ should be $\frac{1}{2}[\delta(\tau+0^+)+\delta(\tau-0^+)]$ when considering operators at $\tau=0$; this introduces some subtleties. Applying standard bosonization techniques \cite{giamarchi2004quantum,Shankar:1995rm} for spin-$1/2$ systems, we map the spin operators to bosonic fields; see Appendix.~\ref{appendix:B} for details. When $\alpha = x$, the most relevant contribution is 
$
    S_d^x = {\mu} \int dxd\tau \delta(\tau) \left(\cos{2\theta_L} + \cos{2\theta_R} + \dots\right)
$
and when $\alpha = z$, it is 
$
    S_d^z = \mu' \int dxd\tau\delta(\tau) (\cos{4\phi_L} + \cos{4\phi_R} +c(\partial_x{\phi_L})^2+c(\partial_x{\phi_R})^2 + \dots)
$
 where $\theta_{L/R}$ and $\phi_{L/R}$ are bosonization variables in the Choi state,  $\mu, \mu'\sim\tanh^{-1}{\left[p/(1-p)\right]}$ with some nonuniversal coefficients from bosonization and $c$ is another unimportant constant from bosonization . To the most relevant order, there is no coupling between the left and right sectors, and near the critical point, the R\'enyi-2 correlator can be factorized into $C^{\text{II}}_{\alpha}(j-i) \sim \langle\hat{\sigma}_{L;i}^\alpha \hat{\sigma}_{L;j}^\alpha\rangle_L\langle\hat{\sigma}_{R;i}^\alpha \hat{\sigma}_{R;j}^\alpha\rangle_R$, where $\langle\cdot\rangle_{L/R}$ is the expectation value of the left and right pieces of the Choi state. The effective field theory of both left and right pieces of the Choi state describes the ground state under weak measurement and post-selection, $\ket{\Psi_0} \rightarrow \hat{K}_\alpha\ket{\Psi_0}$, where the measurement is given by $\hat{K}_\alpha=\prod_{i=1}^L \text{e}^{u\hat{\sigma}^\alpha_i\hat{\sigma}^\alpha_{i+1}/2}$. 

\begin{figure}[htb]
    \centering
        \includegraphics[width=0.65\linewidth]{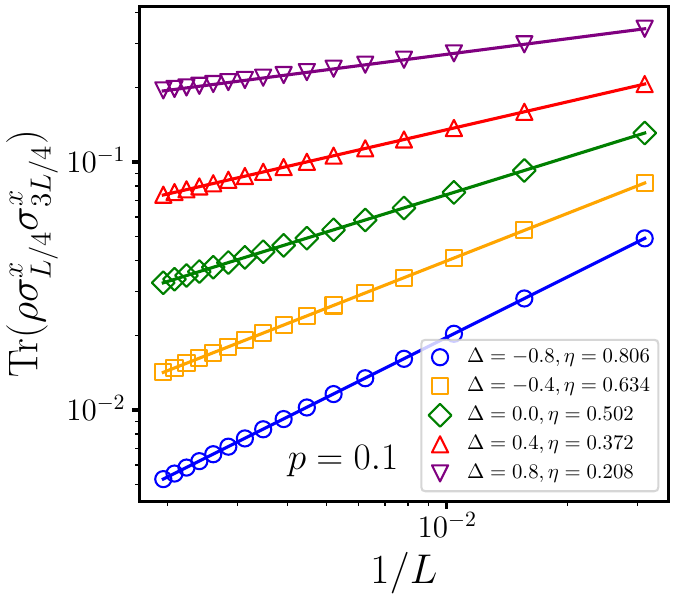}
\caption{Standard correlation function $\Tr[\hat\rho\hat{\sigma}^x_i\hat{\sigma}^x_j]$ at decoherence strength $p=0.1$ for various anisotropy parameters $\Delta$. The numerically extracted critical exponents show excellent agreement with theoretical predictions from Luttinger liquid theory across all parameter values. Simulations were performed for system sizes ranging from $L=32$ to $512$, we demonstrate the result of numerical results (data points) and analytical predictions (solid lines).}
        \label{normal}
\end{figure}

In the standard field theory framework, a phase transition can be understood through the scaling dimensions of operators in the effective Luttinger liquid description \cite{sachdev2011quantum}. By tuning the Luttinger parameter, the scaling dimensions of certain operators can change, rendering them relevant and driving the system toward a new phase. Taking XX decoherence as an example, when decoherence is relevant, a semiclassical solution is such that both $\theta_L$ and $\theta_R$ take values $\pm\pi$ with equal weight. Consequently, the R\'enyi-2 correlations exhibit long-range order and a average ferromagnetic order in the $x$-direction ($\mathrm{xFM}$) emerges:
\begin{equation}
    C^{\text{II}}_{X}(j-i)=|\langle\cos{\theta^i_{L/R}}\cos{\theta^j_{L/R}}\rangle_{L/R}|^2=1
\end{equation}
when $|j-i|\rightarrow\infty$, while the traditional correlation functions remain unmodified by decoherence since $\mathcal{N}^x$ commutes with $\hat{\sigma}^x_i\hat{\sigma}^x_j$:
\begin{equation}
\Tr[\hat\rho\hat{\sigma}^x_i\hat{\sigma}^x_j]=\Tr[\hat{\mathcal{N}}^x[\ket{\Psi_0}\!\bra{\Psi_0}\hat{\sigma}^x_i\hat{\sigma}^x_j]]=\bra{\Psi_0}\hat{\sigma}^x_i\hat{\sigma}^x_j\ket{\Psi_0}.
\end{equation}
Hence, no long range order of strong symmetry emerges, and we also confirm this through numeric (see Fig.~\ref{normal})

Thus, by tuning the parameters, the system undergoes a continuous phase transition that spontaneously break the strong symmetry to a weak one. This transition is characterized by the emergence of long-range R\'enyi-2 correlations, which serves as a clear signature of the SWSSB phase.

\section{The phase diagram under XX decoherence}
\label{section:4}
\subsection{Renormalization group analysis}
\label{section:4.1}
\begin{figure}[tb]
 \centering
    \includegraphics[width=\linewidth]{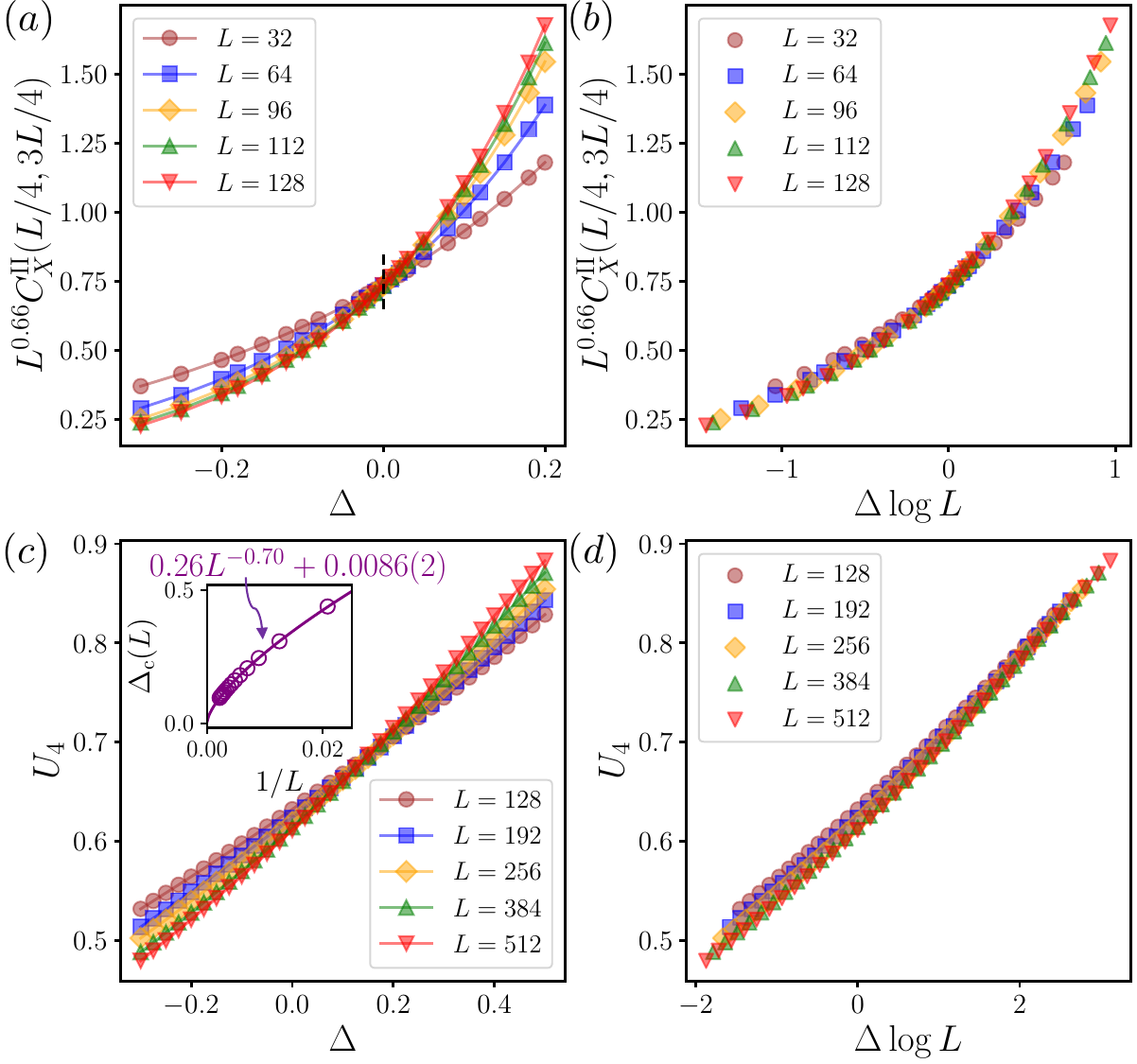}
    \caption{(a)-(b) Direct calculation and data collapse of Rényi-2 correlation function for mixed states and (c)-(d) data collapse of the Binder ratio $U_4 = \frac{1}{2}\left(3 - \frac{\langle \hat O_{\mathrm{xFM}}^4\rangle}{\langle \hat O_{\mathrm{xFM}}^2\rangle^2}\right)$, where $\hat O_{\mathrm{xFM}}=\frac{1}{L}\sum\hat\sigma_i^x$, obtained through effective decoupling theory, i.e., $\ket{\Psi_0}\rightarrow\hat{K}_{X}\ket{\Psi_0}$. $U_{4}$ is a dimensionless physical quantity at the critical point artificially constructed for $\mathbb{Z}_2$ symmetry breaking. Both cases show good agreement with $\Delta_c = 0$, and the $x$-axis is chosen as $\Delta \log L$ in data collapses to be consistent with theoretical expectations. The decoherence strength is $p=0.1$ here. Simulated system size is $L=32$ to $128$ in (a)-(b) and $L=128$ to $512$ in (c)-(d). And we also show the result of data collapse when $p=0.2$ in Appendix.\ref{appendix:F}.}
    \label{collapse_p=0.1.pdf}
\end{figure}

We first consider the XXZ model under two-site XX decoherence, In this entire section, we focus on the XX decoherence; therefore, we omit the subscript in \(\hat{\mathcal{N}}^X\) and simply denote it as \(\hat{\mathcal{N}}\) in this section. The action can be mapped to the standard Kondo problem with a boundary impurity under a space-time rotation. The scaling dimensions of operators are given by $\text{dim}[e^{\ii p\theta}] = \frac{p^2}{4K}$ and $\text{dim}[e^{\ii q\phi}] = \frac{q^2 K}{4}$. From perturbative renormalization group (RG) analysis, the parameter flow is described by the equation $\frac{d\mu}{dl} = (1 - \frac{1}{K})\mu$ for small $\mu$,  $l = \ln\frac{\Lambda_0}{\Lambda}$ is the RG flow parameter, with $\Lambda$ and $\Lambda_0$ representing the running and initial cutoff scales, respectively and $\mu, \sim\tanh^{-1}{\left[p/(1-p)\right]}$ is the effective coupling constant defined in \ref{section:3}. This indicates that the model has a phase transition at $K = 1$ ($\Delta_c = 0$) when $\mu$ is small. For $K < 1$ ($\Delta < 0$), the phase is a trivial Luttinger liquid with strong symmetry, while for $K > 1$ ($\Delta > 0$), the phase exhibits SWSSB. This type of RG equation reveals a boundary BKT-like phase transition with correlation length $\xi \sim e^{\frac{1}{|\Delta - \Delta_c|}}$, which is confirmed by data collapse (see Fig.~\ref{collapse_p=0.1.pdf}). And we remark that the phase transition we discuss arises only in the presence of decoherence without a corresponding counter part at $\Delta_c=0$ without decoherence. The phase transition here couldn't be regarded as a pure state phase transition under decoherence and is thus intrinsic to the open quantum system setting


\begin{figure*}[bt]
    \centering
    \includegraphics[width=0.75\linewidth]{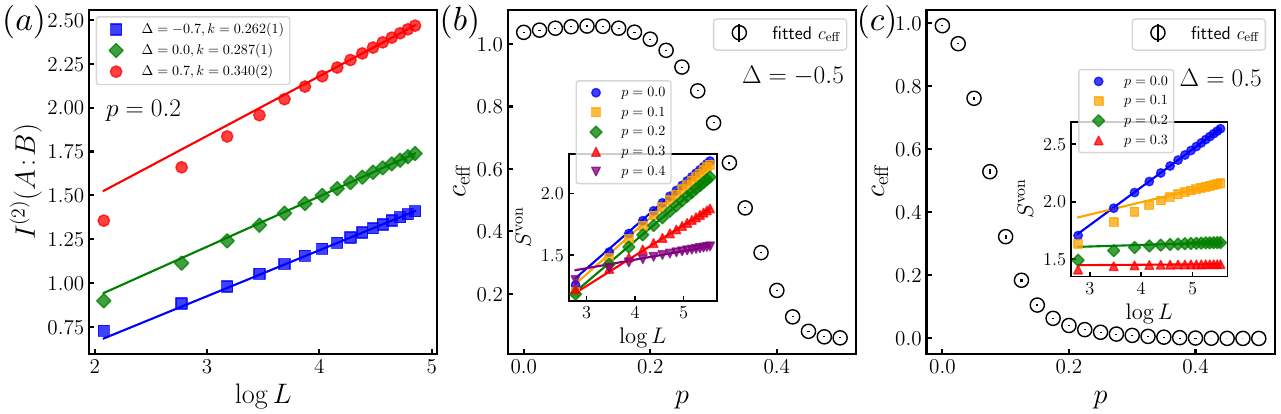}
\caption{(a) Results for the R\'enyi-2 mutual information $I^{(2)}(A:B)$ of the decohered density matrix, where $A$ and $B$ correspond to the left and right halves of the system, respectively. Theoretical predictions give $I^{(n)}(A:B) = 4\dim[\hat{\mathcal{B}}^\alpha_n]\ln{\frac{L}{a}}$ (see the main text for details). For $n=2$, when the decoherence channel is irrelevant, $4\dim[\hat{\mathcal{B}}^{X}_2] = 1/4$, where \(4 \dim[\hat{\mathcal{B}}^{X}_2] = 0.26 \approx \frac{1}{4}\) at \(\Delta = -0.7\);  
otherwise, deviations from \(\frac{1}{4}\) are observed, with  
\(4 \dim[\hat{\mathcal{B}}^{X}_2] = 0.34\) for \(\Delta = 0.7\) and  
\(4 \dim[\hat{\mathcal{B}}^{X}_2] = 0.29\) for \(\Delta = 0\).
 . (b) Direct calculation of the effective central charge $c_{\text{eff}}$ for the Choi state at $\Delta = -0.5$. Here, $c_{\text{eff}} = 1$ remains stable across a wide range of decoherence strengths $p$, while the correlation length $\xi \sim \min[L, \frac{a\mu_0}{\mu}]$, where $\mu_0$ is a constant. Finite-size effects cause deviations from the RG fixed point prediction for large $\mu$. (c) Results for the effective central charge $c_{\text{eff}}$ of the Choi state at $\Delta = 0.5$, showing a rapid decrease of $c_{\text{eff}}$ to zero as $p$ increases. The correlation length follows $\xi \sim \min[L, \frac{a\tilde\mu}{\tilde\mu_0}]$, where $\tilde\mu_0$ is a constant. For small $\mu$, finite-size effects similarly lead to deviations from the RG fixed point prediction. The simulated system size is $L = 8$ to $128$ in (a) and $L = 16$ to $256$ in (b)-(c).}
    \label{entropy}
\end{figure*}

Although the decoherence term $\mu$ is irrelevant in the perturbative regime when $\Delta < 0$, it is still possible for a phase transition to occur at finite $\mu$. This subtlety can be resolved by invoking the strong-coupling/weak-coupling duality between the $\theta$ and $\phi$ fields in the Kondo problem \cite{kane1992transmission,furusaki1993single}, which is also used for measurement-induced phase transitions \cite{PhysRevX.13.021026,sun2023new} after space-time rotation. This duality maps the action $\mathcal{S}^x = \frac{K}{2 \pi }\int dxd\tau\left[(\partial_x\theta)^2+(\partial_\tau\theta)^2\right]+{\mu }\int dx \cos{2\theta}$ to $\tilde{\mathcal{S}}^x = \frac{1}{8 \pi K }\int dxd\tau\left[(\partial_x\phi)^2+(\partial_\tau\phi)^2\right]-\tilde{\mu}\int dx \cos{\phi}$, where the leading order of $\tilde{\mu}$ is given by $\tilde{\mu} = \tilde{\mu}_0 e^{-4{\mu}^{1/2}}$; see also Appendix.~\ref{appendix:C} for details of the duality mapping. Perturbative RG flow using the dual description reads $\frac{d{\mu}}{dl} = 2(K - 1)\sqrt{\mu}$ for large $\mu$. This result shows that the RG flow does not change sign between the weak-decoherence regime ($\mu$ small) and the strong-decoherence regime ($\mu$ large). Thus, it is safe to conclude that there is no phase transition as $p$ increases. 

\subsection{Entanglement properties of phases}
\label{section:4.2}
To verify the field-theory analysis given above from the perspective of entanglement, we first investigate the behaviors of R\'enyi mutual information. The entropy of density matrix can be calculated by standard boundary conformal field theory \cite{calabrese2009entanglement,cardy2006boundary,affleck1997boundary,zou2023prl} yields
$
\Tr\hat\rho_A^n = \langle\langle \Psi^{\otimes n}_0 \otimes \Psi^{*\otimes n}_0 | \hat{\mathcal{N}}_{\alpha}^{*\otimes n}(\hat{S}_n(A)) \hat{\mathbb{I}}^{\otimes n}(\overline{A}) \rangle\rangle,
$
where $\hat{\mathcal{N}}_{\alpha}^{*}$ denotes the complex conjugation of $\hat{\mathcal{N}}_{\alpha}$, $|\Psi^{\otimes n}_0 \otimes \Psi^{*\otimes n}_0 \rangle\rangle=\ket{\Psi_0}^{\otimes n}\otimes\ket{\Psi^*_0}^{\otimes n}$ is written with a slight abuse of notation ,  and $\hat{S}_n(A)$ represents the permutation operator \cite{calabrese2009entanglement} acting on subregion $A=[u,v]$. This calculation maps $\Tr\hat\rho_A^n$ to a two-point correlation function of boundary condition changing operators
$
\Tr\hat\rho_A^n=\langle\hat{\mathcal{B}}^\alpha_n(u)\hat{\mathcal{B}}^\alpha_n(v)\rangle,
$
where $\hat{\mathcal{B}}^\alpha_n$ corresponds to the boundary condition change from $\ket{\hat{\mathbb{I}}^{\otimes n}}\rangle$ to the RG fixed point of $\ket{\hat{\mathcal{N}}^{*\otimes n}_{\alpha}(\hat{S}_n)}\rangle$. The R\'enyi mutual information $I^{(n)}(A:B) =S^{(n)}_A + S^{(n)}_{B} - S^{(n)}_{A \cup B}$ of the left and right half-chain with open boundary condition (OBC) is given by 
\begin{gather}
    I^{(n)}(A:\overline{A})= 4\dim[\hat{\mathcal{B}}^\alpha_n]\ln{\frac{L}{a}},
\end{gather}
where $A=[0,L/2]$. We will use $I(A:B)$ without any superscript to denote the von Neumann mutual information. For irrelevant decoherence channels, the scaling dimension is $\dim[\hat{\mathcal{B}}^\alpha_n]=\frac{c}{24}(n-\frac{1}{n})$, with $c$ being the central charge. Note that for OBC, there is only one set of holomorphic Virasoro algebra, which results in a difference by a factor of 2 compared to the periodic boundary conditions (PBC). When the channel is relevant or marginal, how the conformal boundary condition flow for multi-channel Luttinger liquid remains an open problem, and we even don't know how many boundary states we have for CFT with $c\geq 1$, such as multi-channels Luttinger liquid \cite{Affleck:2000ws,oshikawa2010boundary}. Our numerical observations indicate that the flow leads to another boundary condition changing operator with different scaling dimensions, particularly evident in calculations of $I^{(2)}(A:\overline{A})$ when $\Delta>0$, and we leave the theoretical discussion of this problem to future studies.

Mutual information across all phases exhibits logarithmic scaling as shown in Fig.~\ref{entropy} (a), and even the SWSSB phase does not obey the area law of mutual information, which is typical of traditional SSB, making entanglement-based methods for phase determination complicated. A more direct approach is to calculate the entanglement entropy of the Choi state $|\rho\rangle\rangle$. In the trivial phase, where decoherence is irrelevant, the entanglement entropy follows that of two decoupled Luttinger liquids with central charge $c_{\text{eff}}=1$. The SWSSB phase obeys the area law with central charge $c_{\text{eff}}=0$ (see Figs.~\ref{entropy} (b) and (c)). At criticality, the entanglement structure reveals a novel universality class distinct from free bosonic theories with a variable central charge, as detailed in the following section.

\subsection{Local Recoverability}
\label{section:4.3}
As mentioned in the introduction, there is a fundamental difference between the equivalence of mixed states and pure states. Specifically, any finite-depth local unitary operator can be exactly reversed by its inverse, whereas a finite-depth quantum channel is generally \emph{not} reversible. In this section, we discuss this distinction in greater detail and demonstrate that the SWSSB phase at $(\Delta < \Delta_c, p > 0)$ cannot be recovered to the pure Luttinger liquid phase at $(\Delta, p = 0)$. On the other hand, the mixed-state Luttinger liquid phase with $(\Delta > 0, p > 0)$ can be continuously connected to the pure-state Luttinger liquid at $\Delta = 0$ by employing the concepts of conditional mutual information and the Petz recovery map.

In general, a quantum channel $\hat{\mathcal{N}}$ can be approximately recovered if and only if
\begin{gather}
    S(\rho \| \sigma) \approx S(\hat{\mathcal{N}}(\rho) \| \hat{\mathcal{N}}(\sigma)),
\end{gather}
where $S(\rho \| \sigma) = \operatorname{Tr}[\rho (\log \rho - \log \sigma)]$ is the quantum relative entropy.  
The if part follows from the monotonicity of relative entropy, also known as the data processing inequality,
\begin{gather}
    S(\rho \| \sigma) \geqslant S(\hat{\mathcal{N}}(\rho) \| \hat{\mathcal{N}}(\sigma)) \geqslant S(\hat{\mathcal{R} }\circ \hat{\mathcal{N}}(\rho) \| \hat{\mathcal{R}} \circ \hat{\mathcal{N}}(\sigma)) \approx S(\rho \| \sigma),
\end{gather}
where $\hat{\mathcal{R}}$ is a recovery channel.

And we can prove the only if part by a well-known "pretty good" recovery channel is the \emph{twirled Petz map}, defined as
\begin{align}
   \hat{ \mathcal{R}}_{\sigma, \mathcal{N}}(\cdot)& =\nonumber\\ \int_{-\infty}^{\infty} \mathrm{d}t \, \beta_0(t) \, \sigma^{\frac{1 + i t}{2}} &\, \hat{\mathcal{N}}^\dagger \left[ \hat{\mathcal{N}}(\sigma)^{-\frac{1 + i t}{2}} (\cdot) \, \hat{\mathcal{N}}(\sigma)^{-\frac{1 - i t}{2}} \right] \sigma^{\frac{1 - i t}{2}}
\end{align}
 where the weighting function $\beta_0(t)$ is given by
$
\beta_0(t) = \frac{\pi}{2} \left( \cosh(\pi t) + 1 \right)^{-1}.
$, $\hat{\mathcal{N}}^{\dagger}$ denotes the dual channel of $\hat{\mathcal{N}}_{\alpha}$, defined by $\text{Tr}[\hat{O}^\dagger_1\hat{\mathcal{N}}(\hat{O}_2)] = \text{Tr}[\hat{\mathcal{N}}^{\dagger}(\hat{O}^\dagger_1)\hat{O}_2]$ for any $\hat{O}_{1/2}$,
The twirled Petz map is pretty good in the sense that \begin{equation}
\label{eq14}
    S(\rho \| \sigma) - S(\hat{\mathcal{N}}(\rho) \| \hat{\mathcal{N}}(\sigma)) \geq -2 \log F(\rho, \hat{\mathcal{R}}_{\sigma, \mathcal{N}} \circ \hat{\mathcal{N}}[\rho]),
\end{equation}
where 
$F(\rho, \sigma) := \| \sqrt{\rho} \sqrt{\sigma} \|_1 \leqslant 1$
is the fidelity, and when $S(\rho \| \sigma) = S(\hat{\mathcal{N}}(\rho) \| \hat{\mathcal{N}}(\sigma))$, the fidelity is $1$ and the recovery is exact. However, recovery of the whole decoherence channel is not enough to define many-body quantum phases, since locality plays an important role in quantum many-body systems. To determine whether a quantum channel is locally recoverable, we consider a quantum channel $\hat{\mathcal{N}}_A$ acting on a region $A = [-R/2, R/2]$, and $B$ is a region near the boundary of $A$, where $B = [-r - R/2, R/2 + r]$ and $C$ is the rest of the system. We assume $r \ll R \ll L$, where $L$ is the total system size. We say the channel is locally recoverable if there exists a recovery channel $\mathcal{R}_{AB}$ acting on the region $AB$ that can reverse the effect of $\mathcal{N}_A$.

\begin{figure}[htbp]
    \centering
    \includegraphics[width=0.9\linewidth]{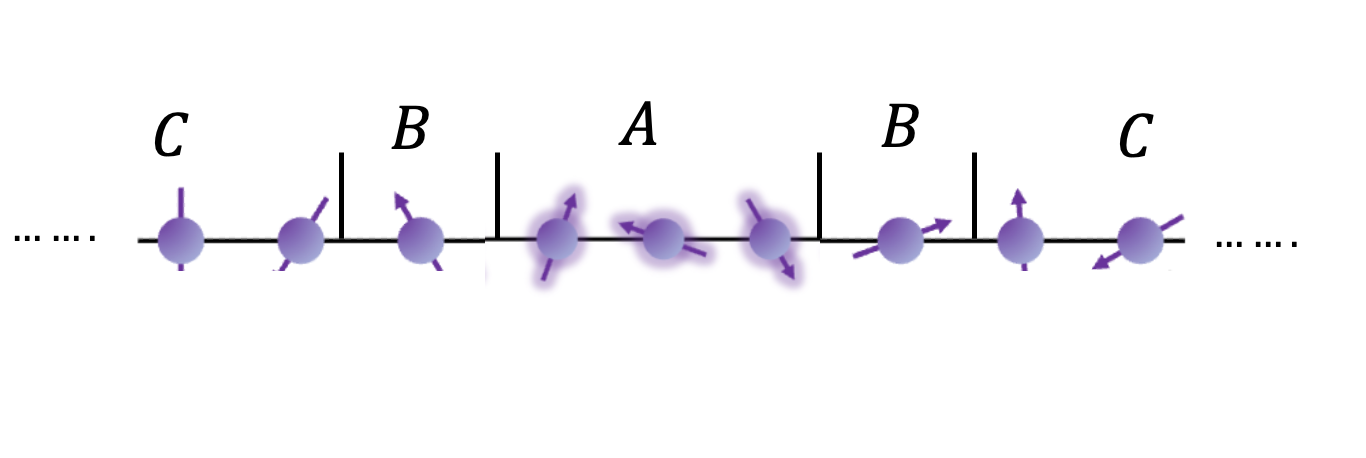}
    \caption{An illustration of the geometry of the system where $A = [-R/2, R/2]$, $B = [-r - R/2, R/2 + r]$, and $C$ is the rest of the system. The quantum channel $\hat{\mathcal{N}}_A$ is the restriction of $\hat{\mathcal{N}}$ to $A$, and the recovery channel $\hat{\mathcal{R}}_{AB}$ acts on $AB$.}
    \label{CMI}
\end{figure}

To realize this recovery, we choose $\sigma = \rho_{AB} \otimes \tilde{\rho}_C$, where $\tilde{\rho}_C$ is a reference state independent of $\rho_{ABC}$. Thus, the twirled Petz map $\hat{\mathcal{R}}_{\sigma, \mathcal{N}_A}(\cdot)$ depends only on $AB$ but not on $C$. Using equation~\eqref{eq14} and the inequality $-\ln F(\rho, \sigma) \geqslant \frac{1}{4 \log 2} \| \rho - \sigma \|_1^2$, we have
\begin{align}
    &I_{\rho}(AB : C) - I_{\hat{\mathcal{N}}(\rho)}(AB : C) \geqslant \nonumber\\ 
    &\quad \frac{1}{2 \log 2} \left\| \hat{\mathcal{R}}_{\sigma, \mathcal{N}_A} \circ \hat{\mathcal{N}}_A(\rho_{ABC}) - \rho_{ABC} \right\|_1^2.
\end{align}
Furthermore, using the conditional mutual information $I(A : C | B) = I(AB : C) - I(B : C)$, we can rewrite
\begin{gather}
    I_{\rho}(AB : C) - I_{\hat{\mathcal{N}}_A(\rho)}(AB : C)\nonumber\\ = I_{\rho}(A : C | B) - I_{\hat{\mathcal{N}}(\rho)}(A : C | B).
\end{gather}
Since for gapped quantum systems $I(A : C | B)$ decreases exponentially with the size of $B$, this provides a quantitative measure of local recoverability.

Below, we choose $\rho = \ket{\Psi_0}\bra{\Psi_0}$, the ground state of the $XXZ$ Hamiltonian, and the Rényi-$n$ mutual information $I^n_{\mathcal{N}_A(\ket{\Psi_0}\bra{\Psi_0})}(AB : C)$ can be calculated by a four-point correlation function in CFT:
\begin{align}
    I^n_{\hat{\mathcal{N}}_A(\ket{\Psi_0}\bra{\Psi_0})}(AB : C) =\nonumber\\ -\frac{2}{n-1} \log \langle \hat{\mathcal{C}_1}(-R/2 - r) &\hat{\mathcal{C}}_2(-R/2) \hat{\mathcal{C}}_1^\dagger(R/2) \hat{\mathcal{C}}_2^\dagger(R/2 + r) \rangle,
\end{align}
where $\hat{\mathcal{C}}_1$ is the boundary condition changing operator from $\ket{\hat{\mathbb{I}}}$ to $\ket{\hat{S}_n}$  , and $\hat{\mathcal{C}}_2$ is the boundary condition changing operator from $\ket{\hat{S}_n}$ to $\ket{\hat{\mathcal{N}}^{*\otimes n}(\hat{S}_n)}$. 

As discussed and numerically verified in the previous section, when $\hat{\mathcal{N}}$ does not drive the system into the SWSSB phase, the channel is irrelevant and $\ket{\hat{\mathcal{N}}^{*\otimes n}(\hat{S}_n)}$ flows to $\ket{\hat{S}_n}$ under renormalization, and $\hat{\mathcal{C}}_2$ becomes the identity and  operator and we assume \begin{gather}
\langle \hat{\mathcal{C}}_1(-R/2 - r) \hat{\mathcal{C}}_2(-R/2) \hat{\mathcal{C}}_2^\dagger(R/2) \hat{\mathcal{C}}_1^\dagger(R/2 + r) \rangle\nonumber\\=\langle\hat{\mathcal{C}_1}(-R/2 - r)  \hat{\mathcal{C}}_1^\dagger(R/2 + r) \rangle+\lambda f(r,R)R^{-h}
\end{gather}
where $f(r,R)$ is a continuous function of order $O(1)$ and the second term comes from the deviation of $\ket{\hat{\mathcal{N}}^{*\otimes n}(\hat{S}_n)}$ and real conformal fixed point at finite size. Consequently,
\begin{gather}
    I^n_{\ket{\Psi_0}\bra{\Psi_0}}(AB : C) - I^n_{\hat{\mathcal{N}}_A(\ket{\Psi_0}\bra{\Psi_0})}(AB : C) = O(R^{-h_\Delta}\log R),
\end{gather}
where $h_\Delta=(h-2\dim[\hat{\mathcal{C}}_n])>0$ so there exists an approximately local recovery map from $\hat{\mathcal{N}}_A(\ket{\Psi_0}\bra{\Psi_0})$ back to $\ket{\Psi_0}\bra{\Psi_0}$.

On the other hand, when the channel is relevant and drives the state into the SWSSB phase, we can calculate the OPE of $\hat{\mathcal{C}}_1$ and $\hat{\mathcal{C}}_2$, where $\hat{\mathcal{C}}_1(0)\hat{\mathcal{C}}_2(r)\sim r^{-h_\delta}\hat{C}_{12}(0)$ and $\dim[\hat{C}_{12}]=h_\gamma$. In general, when $r \ll R$, we have
\begin{gather}
    I^n_{\hat{\mathcal{N}}_A(\ket{\Psi_0}\bra{\Psi_0})}(AB : C) \sim -\log {r^{-2h_\delta}}{R^{-h_{\gamma}}}.
\end{gather}
As a result,
\begin{gather}
    I^n_{\ket{\Psi_0}\bra{\Psi_0}}(A : C | B) - I^n_{\hat{\mathcal{N}}(\rho)}(A : C | B) = c_a \log R - c_b \log r,
\end{gather}
where $c_a$ and $c_b$ are universal constants depending on the boundary conditions. However, as noted in the previous section, even the classification of conformal boundary states for multi-channel Luttinger liquids remains an open problem, so we cannot provide more theoretical information on these constants. Moreover, the channel cannot be recovered locally since $r$ needs to scale with $R$ to achieve good recovery.

We remark that our calculation can be applied to any Rényi-$n$ index, and in general it will give different phase boundaries. The von Neumann conditional mutual information precisely captures whether a channel can be locally recovered, which generally differs from the phase boundary given by the Rényi-2 like quantities. However, our work mainly focuses on the Rényi-2 quantity, and as we have shown in previous discussions, Rényi-2 quantities can serve as order parameters for the symmetry breaking $G_L \times G_R \to G_{\mathrm{diag}}$ in the Choi state, where $G_\text{diag}$ is diagonal subgroup of $G_L \times G_R \to G_{\mathrm{diag}}$. Compared to Rényi-1-like quantities such as fidelity correlators, Rényi-2 has a more direct relation to symmetry breaking. We aim to discuss the recoverability in terms of the Rényi-2 conditional mutual information difference,
which can be understood through the claim in \cite{zhang2024strong}: two states are equivalent if and only if they are connected by a finite Lindbladian evolution that maintains analytic variation, provided that 
$
    I^2_{\ket{\Psi_0}\bra{\Psi_0}}(A : C | B) - I^2_{\hat{\mathcal{N}}(\rho)}(A : C | B)
$
is small. Moreover, the phase boundary of recoverability defined by analytic finite Lindbladian evolution coincides with the discussion presented in the previous sections.

\subsection{Physics on the critical line}
\label{section:4.4}
\begin{figure}[thbp]
    \centering
    \includegraphics[width=0.9\linewidth]{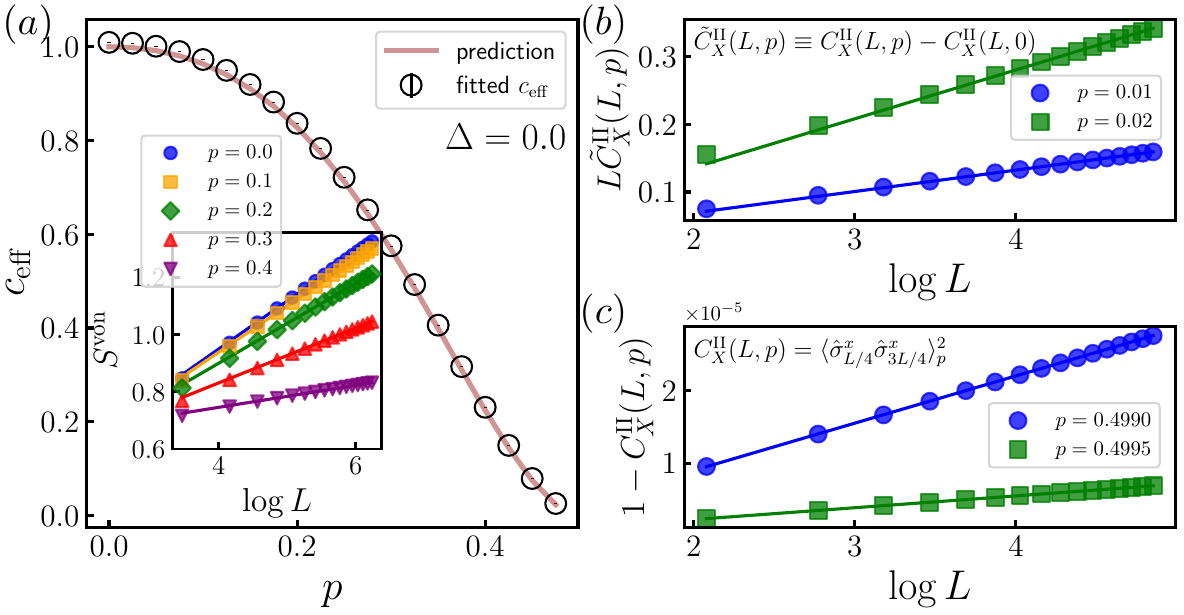}
    \caption{(a) The effective central charge $c_{\text{eff}}$ from numerical calculations of half-chain entanglement entropy for $\hat{K}_X \ket{\Psi_0}$ at $\Delta_c = 0$, comparing with the theoretical prediction~\eqref{eq:ceff} (the pink solid line). The inset shows the fitting of $S^\text{von}(L/2) \sim \frac{c_\text{eff}}{6}\ln L$ for different $p$. (b)-(c) verify the perturbation calculation of R\'enyi-2 correlation function in the strong and weak decoherence limits at $\Delta_{c} = 0$ (see Eqs.~\eqref{small_p} and \eqref{large_p}). Simulated system size is $L=32$ to $512$ in (a) and $L=8$ to $128$ in (b)-(c).}
    \label{varyc}
\end{figure} 
\begin{figure*}[hbpt]
    \centering
    \includegraphics[width=0.75\linewidth]{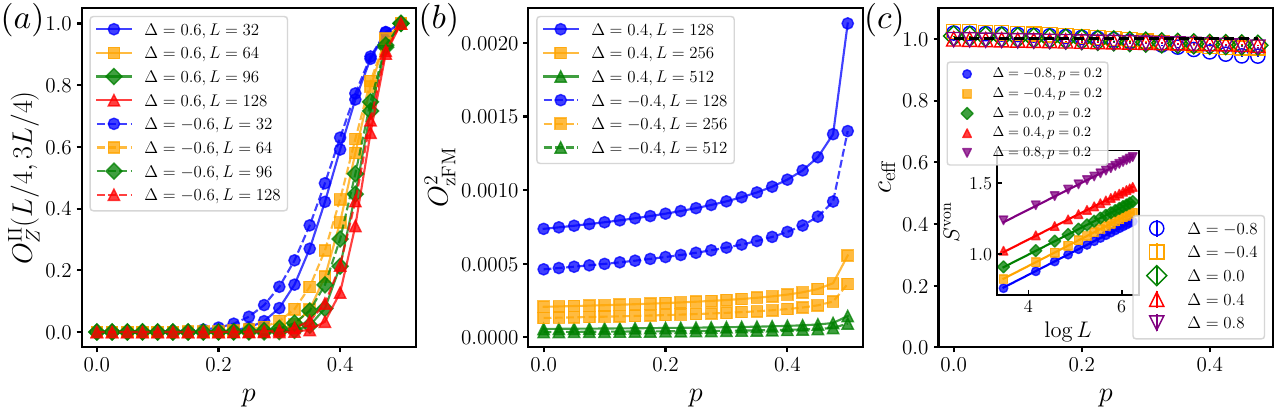}
    \caption{(a) Direct calculation of the Rényi-2 ZZ correlator for the Choi state. As the system size increases, the correlation function rapidly approaches zero. (b) The squared order parameter $O_\text{zFM}^{2} = \frac{1}{L^2} \langle( \sum_{i} \hat{\sigma}_{i}^{z} )^{2}\rangle$ calculated from the effective description under weak measurement $\hat{K}_Z\ket{\Psi_0}$. $O_\text{zFM}^{2}$ remains close to zero for various $p$. (c) Effective central charge $c_{\text{eff}}$ of $\hat{K}_Z\ket{\Psi_0}$ under different parameters, consistently remaining close to 1, showing no deviation from the pure-state Luttinger liquid. Simulated system size is $L=32$ to $512$ in (c).}
    \label{ap_ZZ}
\end{figure*} 
At the critical line ($\Delta_c = 0$), decoherence becomes marginal, enabling us to focus on the most relevant operators. This presents a unique opportunity for analytical investigation. Furthermore, to overcome the substantial finite-size effects at criticality, we mainly study the effective pure state description given by $\hat{K}_X\ket{\Psi_0}$ rather than the Choi state itself numerically. This approach offers superior numerical stability and enables large-scale numerical computations. The scale invariance at criticality dictates that all correlation functions must take the form of scaling functions. For the weak decoherence regime ($p \ll 1$), standard perturbation theory reveals the scaling behavior of R\'enyi-2 correlator which is numerically confirmed in Fig.~\ref{varyc} (b), and details of perturbation could be found in Appendix.~\ref{appendix:D1}:
\begin{gather}
    C^{\text{II}}_{X}(x_2 - x_1) \sim \frac{1}{|x_2 - x_1|} \left( 1 + 2\mu C_1 \ln \frac{|x_2 - x_1|}{a} \right) \, ,
    \label{small_p}
\end{gather}
where $a$ represents the short-distance cutoff and $C_1$ is a non-universal constant. In the complementary strong decoherence regime ($p \to 1/2$), we leverage the powerful strong-weak coupling duality. By mapping the vortex operator to a non-local operator and then applying perturbation theory in the dual picture, we obtain ( details of perturbation could be found in Appendix.~\ref{appendix:D2}):
\begin{gather}
\label{large_p}
    C^{\text{II}}_{X}(x_2 - x_1) \sim 1 - 2\tilde\mu^2 C_2 \ln \frac{|x_2 - x_1|}{a} \, ,
\end{gather}
with $C_2$ being another non-universal constant. This scaling behavior is also verified via simulations as seen in Fig.~\ref{varyc} (c).

A remarkable consequence of the effective decoupling of the Choi state is that we can perform a precise analytical calculation of the half-chain entanglement entropy of the Choi state $|\rho\rangle\rangle$. Through conformal transformation, this calculation maps to determining the entanglement entropy of two independent systems with a boundary defect $\hat{H}_d = -\mu \cos(2\hat{\theta}(0))$ at their center \cite{yang2023entanglement,sun2023new}, see also Appendix.~\ref{appendix:D3} for details. Furthermore, since the XX model realizes a free fermion theory at criticality, we can rewrite our theory in terms of fermionic fields:
\begin{align}
    \hat{H} = \frac{1}{2}\int_{-L}^{L} dx \, \hat{\Psi}^\dagger(x) (\hat{\mathbb{I}} \otimes \hat{\sigma}_z) (-\ii \partial_x )\hat{\Psi}(x) \nonumber \\  
    + \frac{u}{2} \delta(x)\hat{\Psi}^\dagger(x) (\hat{\sigma}_y \otimes \hat{\sigma}_y) \hat{\Psi}(x) \, .
\end{align}
Here, $\hat{\Psi} = (\hat{c}_L, \hat{c}_R, \hat{c}^\dagger_L, \hat{c}^\dagger_R)^T$ is the Nambu spinor, and $\hat{c}_{L/R}$ represents left/right moving chiral fermions. This Hamiltonian describes a gapless Dirac theory with a mass defect,  equivalent to two Ising conformal field theories with a mass defect. The entanglement entropy of free fermions has been calculated analytically for arbitrary defect strength \cite{eisler2010entanglement,Brehm:2015lja,oshikawa1997boundary}, the half chain entanglement of choi state $|\rho\rangle\rangle$, $S^{\text{von}}(L/2) \sim \frac{c_{\text{eff}}}{3} \ln \frac{L}{a}$, yielding an effective central charge:
\begin{align}
    c_{\text{eff}} = -\frac{6}{\pi^2} \Big\{ &[(1+s) \log(1+s) + (1-s) \log(1-s)] \log(s) \nonumber \\
    & + (1+s) \text{Li}_2(-s) + (1-s) \text{Li}_2(s) \Big\} \, ,
    \label{eq:ceff}
\end{align}
where $s = 1/\cosh(2u)$ characterizes the defect strength, and $\text{Li}_2(x) = -\int_0^x dx' \frac{\ln(1-x')}{x'}$. In particular, as shown in Fig.~\ref{varyc}~(a), the $c_\text{eff}$ obtained in our simulations agrees well with the prediction of our theory.

\section{The effect of ZZ decoherence}
\label{section:5}
In this section, we investigate the effect of ZZ decoherence.As mentioned in Section.\ref{section:3} the effective field theory for ZZ decoherence is given by 
\begin{equation}
 S_d^z = \mu' \int dxd\tau\delta(\tau) \big[\cos{4\phi_L} + \cos{4\phi_R} +c(\partial_x{\phi_L})^2+c(\partial_x{\phi_R})^2 \big]
\end{equation}

RG results indicate that the operator $\cos\dim [4\phi_{L/R}]=4K$ becomes relevant when $K<1/4$, and remains irrelevant throughout the regime $|\Delta| < 1$ in critical XXZ model and $\text{dim}[(\partial_x\phi_{L/R})^2]=2$ is also irrelevant, thereby precluding the possibility of SWSSB phase transitions within this interval. While previous studies suggested the existence of a system-environment entanglement transition in the XXZ chain under ZZ decoherence at finite $p$ when $\Delta$ is fixed~\cite{Kuno:2025mcr} through $g$-function, both our numerical results and perturbative field theory show no evidence of phase transitions in the R\'enyi-2 ZZ correlator, as shown in Fig.~\ref{ap_ZZ}. We note that, in the case of single-site Z decoherence, it has been reported that the $g$-function can exhibit a transition at a finite $p$ due to non-perturbative effects associated with the kinetic term $(\partial_x\phi_{L/R})^2$~\cite{PhysRevB.110.094404,PhysRevB.80.184421}. Moreover, the effective central charge of Choi state consistently maintains $c_{\text{eff}}=1$ (see Fig.~\ref{ap_ZZ} (c)), which matches that of a Luttinger liquid without decoherence, and we conclude that it's safe to only consider the vortex like interaction after bosonization.

\section{Generic Decoherence Channels}
\label{section:gen}
While our analysis has primarily focused on XX and ZZ decoherence channels, our theoretical framework naturally extends to more general decoherence patterns that preserve strong symmetries. As a concrete example, consider the most general two-site Pauli decoherence channel:

\begin{gather}
    \hat{\mathcal{N}}_i:\hat\rho\rightarrow(1-p_x-p_y-p_z)\hat\rho+\sum_{\alpha={x,y,z}}p_\alpha \hat{\sigma}_i^\alpha\hat{\sigma}_{i+1}^\alpha\rho\hat{\sigma}_i^\alpha\hat{\sigma}_{i+1}^\alpha
\end{gather}

Within the doubled Hilbert space formalism, any completely positive quantum channel admits an exponential representation:

\begin{gather}
    \ket{\rho}\rangle\rightarrow \exp{(-\sum_{j}\hat{k}^{j}_L\otimes \hat k^{*,j}_R)}\ket{\rho}\rangle
\end{gather}
where $\hat{k}^{j}_L$ and $\hat k^{*,j}_R$ constitute a set of strong symmetry-preserving local operators. Through bosonization, these operators map to specific boundary interactions in the field theory description. 

This framework remains applicable even for multi-site decoherence channels. We do not expect there should be qualitative differences, since for each quantum trajectory, such a noise action can be decomposed into local quantum channel in $O(1)$ depth strong symmetry preserving local unitaries, under which both Rényi-2 correlation and quantum entanglement are stable under short range unitaries. For instance, the leading term of the four-sites XXXX decoherence is still $\cos(2\theta_{L/R})$ and the leading vortex term of the four- sites ZZZZ decoherence is still $\cos(4\phi_{L/R})$, so the behavior is qualitatively similar to our discussion of 2-sites decoherence.

From the perspective of effective field theory, the general nontrivial actions of the three $\mathbb{Z}_2$ symmetries on the Luttinger liquid variables takes the form $-\phi_{L/R}\rightarrow \phi_{L/R}+\frac{\pi}{2}$, $\theta_{L/R}\rightarrow\theta_{L/R}+\pi$ and $\theta_{L/R}\rightarrow-\theta_{L/R}$, and a general action is given by either 
$
    \mathcal{S}^n_{L/R} = \frac{K}{2 \pi }\int dxd\tau\left[(\partial_x\theta_{L/R})^2+(\partial_\tau\theta_{L/R})^2\right]+{\mu }_n\int dx \cos{(2n\theta_{L/R})}
$, if the decoherence is local in $\theta$ variables or 
$\mathcal{S}^m_{L/R} = \frac{1}{2 \pi K }\int dxd\tau\left[(\partial_x\phi_{L/R})^2+(\partial_\tau\phi_{L/R})^2\right]+{\mu }_m\int dx \cos{(4m\phi_{L/R})}$ if the decoherence is local in $\phi$ variables. A similar SWSSB phase transition belonging to the boundary BKT universality class can be identified at critical Luttinger parameters $K_c = n^2$ for $\mathcal{S}^n_{L/R}$ and $K_c = \frac{1}{4m^2}$ for $\mathcal{S}^m_{L/R}$, where the system exhibits analogous critical scaling behavior and universal properties similar with our previous discussion. The duality can also be generalized from $2\theta_{L/R}\leftrightarrow \phi_{L/R}$ to $2k\theta_{L/R}\leftrightarrow \frac{1}{k}\phi_{L/R}$, where $k\in\mathbb{Z}$ and $K\leftrightarrow 4K$. 

This unified description provides a powerful tool for analyzing the phase diagram and critical properties of one-dimensional quantum systems - including bosonic systems, spin-1/2 chains and fermionic systems - whose low-energy effective theory can be described by Luttinger liquid theory, under arbitrary symmetry-preserving decoherence channels.

\section{Conclusions and discussions.}
\label{section:6}

To conclude, we uncover a novel quantum SWSSB transition in the critical XXZ spin chain under local two-site XX decoherence. Unlike conventional SWSSB driven by decoherence, which resembles a thermal phase transition, in our case, the quantum SWSSB is driven solely by the Hamiltonian parameter $\Delta$ and persists for arbitrary decoherence strength, belonging to a boundary BKT universality class with a varying effective central charge, as supported by comprehensive field-theoretical analysis and numerical simulations. The SWSSB in our model occurs with arbitrary coupling to the environment and is purely quantum in nature, similar to quantum phase transitions that can be tuned by adjusting Hamiltonian parameters instead of temperarure. Additionally, no SWSSB transition is observed under the ZZ decoherence channel in critical XXZ spin chain with $|\Delta|<1$, which confirms our analysis by a negative result. The framework build in our paper can be readily generalized to other one-dimensional critical systems. Our findings reveal a distinctive phenomenon with no direct counterpart in traditional spontaneous symmetry breaking. Our work provides a general mechanism for realizing quantum SWSSB in one-dimensional quantum critical states under generic symmetric local decoherence, opening a new avenue for exploring symmetry breaking in mixed-state quantum phases of matter.


We further discuss the experimental relevance of our findings. Efficient preparation of critical states, such as the ground state of the XXZ model, is typically achieved using a quantum circuit \cite{VanDyke:2021nuz,raveh2024deterministic}. After state preparation, the dynamics of decoherence can be described by a Lindbladian framework without a unitary Hamiltonian:
\begin{equation}
\partial_t \hat{\rho}(t) = \sum_i \hat{L}_i \hat{\rho}(t) \hat{L}_i^\dagger - \frac{1}{2} \left( \hat{L}_i^\dagger \hat{L}_i \hat{\rho}(t) + \hat{\rho}(t) \hat{L}_i^\dagger \hat{L}_i \right).
\end{equation}
When $\hat{L}_i = \hat{\sigma}^\alpha_i \hat{\sigma}^\alpha_{i+1}$, this yields $\hat{\rho}(t) = \hat{\mathcal{N}}^{\alpha}[\hat{\rho}(0)]$ with $t = \tanh^{-1}\left[p/(1-p)\right]$. This precisely aligns with our discussion in the paper, indicating that even when environmental noise maintains strong symmetry, the strong symmetry of the XXZ ground state remains unstable in the parameter range $\Delta > 0$.

Another notable consideration is that R\'enyi-2 correlators are not directly measurable in experiments. Reference \cite{sun2024scheme} proposes a concrete protocol for their detection by generalizing the method of measuring R\'enyi entropy through randomized measurements \cite{elben2023randomized,Brydges2019}. This protocol entails performing random Pauli measurements on both the original quantum state and the state after evolution with charged operators, making experimental detection of strong-to-weak symmetry breaking feasible on near-future quantum devices \cite{preskill2018quantum,endo2021hybrid,arute2019quantum,kandala2019error,bharti2022noisy}.

\begin{acknowledgments}

We thank Yuto ASHIDA, Zongping GONG, Masahito YAMAZAKI, Zijian WANG, and Hai-Qing LIN for helpful discussion. Numerical simulations based on matrix product states~\cite{white1992prl,white1993prb,cirac2006prb}, where details could be found in Appendix.~\ref{appexdix:E}, were carried out with the ITENSOR \verb|C++| package~\cite{itensor}. X.-J. Yu was supported by the National Natural Science Foundation of China (Grant No.12405034) and a start-up grant from Fuzhou University. Y.G. is financially supported by the Global Science Graduate Course (GSGC) program at the University of Tokyo. 
The work of S.Y. is supported by China Postdoctoral Science Foundation (Certificate Number: 2024M752760).
\end{acknowledgments}


\appendix
\onecolumngrid

\section{More on Strong and Weak Symmetries and Strong to weak symmetry breaking}

\subsection{Symmetry of quantum channel}
\label{appendix:A.1}
While the main text introduce the notions of symmetry for mixed states, here we discuss the symmetry classification of quantum channels. This distinction is crucial for understanding what is a SWSSB. We consider quantum channels given by completely positive trace preserving (CPTP) maps with symmetry group $G$, where the channel action is described by Kraus operators $\{\hat K_{\alpha}\}$ transforming a state $\hat\rho$ as $\hat\rho \rightarrow \sum_{\alpha} \hat K_{\alpha} \hat \rho \hat K_{\alpha}^{\dagger}$.

\begin{itemize}
    \item \textbf{Strong symmetry:} A channel exhibits strong symmetry when each Kraus operator individually commutes with the symmetry operations:
    \begin{equation}
        \hat{U}_g \hat K_{\alpha} = \hat K_{\alpha} \hat{U}_g \quad \forall \alpha, \forall g \in G
    \end{equation}
    This ensures symmetry preservation in every quantum trajectory. 
    \item \textbf{Weak symmetry:} The channel possesses weak symmetry when the combined action of all Kraus operators respects the symmetry, even if individual Kraus operators do not. Mathematically:
    \begin{equation}
        \hat{U}_g \left( \sum_{\alpha}\hat K_{\alpha} \hat\rho \hat K_{\alpha}^{\dagger} \right) \hat U^{\dagger}_g = \sum_{\alpha}\hat K_{\alpha} \hat\rho \hat K_{\alpha}^{\dagger} \quad \forall g \in G
    \end{equation}
    In the basis of Kraus operators, this condition allows for:
    \begin{equation}
        \hat U_g \hat K_\alpha = e^{\ii\theta_\alpha(g)}\hat K_\alpha \hat U _g
    \end{equation}
    where the phase factors $e^{\ii\theta_\alpha(g)}$ cannot be eliminated through gauge transformations of $U(g)$. 
\end{itemize}

In the main text, the Kraus operator of our quantum channel is $\hat K^i_1=\sqrt{(1-p)}\hat{\mathbb{I}}_i\otimes\hat{\mathbb{I}}_{i+1}, K^i_2=\sqrt{p}\hat{\sigma}^{\alpha}_i\hat{\sigma}^{\alpha}_{i+1}$, under $\mathbb{Z}_2^Y$ symmetry $\hat{U}_Y=\prod_{i=1}^L\hat\sigma_i^y$. We have $\hat{K}_1\hat{U}_Y=\hat{U}_Y\hat{K}_1$ and $\hat{K}_2\hat{U}_Y=\hat{U}_Y\hat{K}_2$. As a result, the two-site decoherence channels satisfy the strong symmetry condition, since it commutes with all three $\prod_{i=1}^L\hat\sigma_i^{x/y/z}$. On the other hand, if we consider single site case $\tilde{\hat K}_1^i=\sqrt{(1-p)}\hat{\mathbb{I}}_i$ and $\tilde{\hat K}_2^i=\sqrt{p}\hat{\sigma}^{\alpha}_i$, we have $\tilde{\hat{K}}_1\hat{U}_Y=\hat{U}_Y\tilde{\hat{K}}_1$ and $\tilde{\hat{K}}_2\hat{U}_Y=-\hat{U}_Y\tilde{\hat{K}}_2$, with only weak symmetry. And this is why we need to consider two-site decoherence to show SWSSB. This classification explains the key phenomenon discussed in the main text: when a state exhibiting strong symmetry evolves under a channel also with strong symmetry, the resulting state can only maintain weak symmetry. 

\subsection{Correlators used for SWSSB}
\label{appendix:A.2}
First, let us briefly review why the R\'enyi-2 correlator can effectively capture the transition from strong to weak symmetry breaking using the Choi state. We consider the following expression:
\begin{gather}
    C^{\text{II}}_{O}(i-j) = \frac{\langle\langle\rho|\hat{O}^{\alpha}_{L;i}\hat{O}^{*,\alpha}_{R;i}\hat{O}^\dagger_{L;j}\hat{O}^{\alpha,T}_{R;j}|\rho\rangle\rangle}{\langle\langle\rho|\rho\rangle\rangle}.
\end{gather}
The operator \(\hat{O}^\alpha\) transforms under a representation of the symmetry group \(G\) as $
\hat{U}_g^\dagger \hat{O}^\alpha \hat{U}_g = U_g^{\alpha\beta} \hat{O}^\beta$. 
In the presence of strong symmetry \(G\), the state transforms as 
$
|\rho\rangle\rangle \rightarrow \hat{U}_{g;L} |\rho\rangle\rangle,
$
indicating that the operator \(\hat{O}^\alpha\) is charged. It can only yield non-zero values when symmetry is broken. On the other hand, under weak symmetry breaking, the state transforms as $
|\rho\rangle\rangle \rightarrow \hat{U}_{g;L} \hat{U}^*_{g;R} |\rho\rangle\rangle.$
The transformation of the operators is given by 
\begin{gather}
\hat{O}^{\alpha}_{L;i} \hat{O}^{*,\alpha}_{R;i} \rightarrow U_{\alpha\beta} U_{\alpha\gamma}^* \hat{O}^{\beta}_{L;i} \hat{O}^{*,\gamma}_{R;i} = \hat{O}^{\alpha}_{L;i} \hat{O}^{*,\alpha}_{R;i}.
\end{gather}
In this case, the operators remain in a trivial representation under the symmetry, allowing them to yield non-zero values even when weak symmetry is present.

Although we focus on Rényi-2 correlator in the main text, there are also some other choices to diagnose SWSSB. In Ref.~\cite{lessa2024strong,weinstein2024efficient,liu2024diagnosing}, the authors proposed that the fidelity correlator serves as a useful diagnostic tool for detecting strong-to-weak spontaneous symmetry breaking (SW-SSB). For a mixed state $\rho$ and a pair of local operators $O(x)$ and $O(y)$ carrying symmetry charge, the fidelity correlator is defined as:
\begin{equation}
    F_O(x,y) = F(\rho, O(x)O^\dagger(y)\rho O(y)O^\dagger(x)),
\end{equation}
where $F(\rho,\sigma)=\Tr\sqrt{\sqrt{\rho}\sigma\sqrt{\rho}}$ is the fidelity between two density matrices. When $\rho$ has strong symmetry under group $G$, we have $U(g)\rho U^\dagger(g)=\rho$ for all $g\in G$. The fidelity correlator captures the distinguishability between the original state $\rho$ and the state after local symmetry operations.

For a state with SW-SSB, the fidelity correlator exhibits long-range order:
\begin{equation}
    \lim_{|x-y|\to\infty} F_O(x,y) = c > 0,
\end{equation}
while conventional correlation functions decay to zero:
\begin{equation}
    \lim_{|x-y|\to\infty} \Tr[\rho O(x)O^\dagger(y)] = 0.
\end{equation}

The fidelity correlator's effectiveness in detecting SW-SSB stems from its ability to capture the quantum state distinguishability that persists at large distances. However, the square root operation on density matrices introduces significant computational complexity for both numerical and theoretical calculations. This restricts the exact calculation of fidelity correlators to a limited class of analytically solvable models. In contrast, the Rényi-2 correlator discussed in the main text avoids these computational challenges while still providing a reliable diagnostic tool for SWSSB phases.

Another diagnostic tool for SWSSB is the Rényi-1 correlator. Unlike the Rényi-2 correlator which utilizes the Choi state representation, the Rényi-1 correlator is defined through the canonical purification of a mixed state $\rho$. For local operators $O(x)$ and $O(y)$ carrying symmetry charge, it takes the form:
\begin{equation}
    R_1(x,y) = \Tr[O(x)O^\dagger(y)\sqrt{\rho}O(y)O^\dagger(x)\sqrt{\rho}]
    = \langle\langle\sqrt{\rho}|O^L_xO^{L\dagger}_yO^R_yO^{R\dagger}_x|\sqrt{\rho}\rangle\rangle,
\end{equation}
where $|\sqrt{\rho}\rangle\rangle$ denotes the canonical purification of $\rho$, with the double-bracket notation emphasizing states in the doubled Hilbert space. The superscripts $L,R$ indicate operators acting on the left and right subsystems respectively. 

A state exhibits SWSSB when the Rényi-1 correlator maintains long-range order:
\begin{equation}
    \lim_{|x-y|\to\infty} R_1(x,y) = c > 0,
\end{equation}
while conventional correlation functions decay to zero:
\begin{equation}
    \lim_{|x-y|\to\infty} \Tr[\rho O(x)O^\dagger(y)] = 0.
\end{equation}
The physical interpretation of the Rényi-1 correlator is particularly transparent: it measures the persistence of correlations in the purified state even when the original mixed state shows no long-range order. This diagnostic tool combines theoretical advantages with experimental accessibility, as measuring $R_1(x,y)$ reduces to measuring standard correlation functions in the purified system if one can prepare the canonical purification state.

\section{Effective Field theory of Choi state}
\label{appendix:B}
The one-dimensional XXZ model provides a paradigmatic example of how a spin system can be mapped to interacting fermions and subsequently to a Luttinger liquid. The Hamiltonian reads:
\begin{equation}
    \hat{H}_{XXZ}=-J\sum_{i}^L (\hat{\sigma}^x_i\hat{\sigma}^x_{i+1}+\hat{\sigma}^y_i\hat{\sigma}^y_{i+1}+\Delta\hat{\sigma}^z_i\hat{\sigma}^z_{i+1}),
\end{equation}
where $\Delta$ characterizes the anisotropy in the $z$ direction.

The first step is to map the spin operators to fermionic operators via the Jordan-Wigner transformation:
\begin{align}
    \hat\sigma^i_z &= 1-2\hat{c}_i^\dagger\hat{c}_i \\
    \hat\sigma^i_+ &= (-1)^i\prod_{j<i}(1-2\hat{c}_j^\dagger\hat{c}_j)\hat c^\dagger_i 
\end{align}
where $c_j$ and $c^\dagger_j$ are fermionic operators satisfying canonical anticommutation relations. The Hamiltonian becomes:
\begin{equation}
    H = 2{J}\sum_j (c^\dagger_j c_{j+1} + c^\dagger_{j+1} c_j) + 4J\Delta\sum_j (n_j-\frac{1}{2})(n_{j+1}-\frac{1}{2}),
\end{equation}
where $n_j = c^\dagger_j c_j$ is the fermion number operator.

To proceed to the Luttinger liquid description, we first express the fermionic operators in terms of a smooth field $\psi(x)$ near the Fermi points $\pm k_F$:
\begin{equation}
    \hat{c}_j \approx \sqrt{a}[\hat\psi_R(x)e^{ik_Fx} + \hat\psi_L(x)e^{-ik_Fx}],
\end{equation}
where $a$ is the lattice spacing, and $\psi_{R/L}$ are the right/left-moving components. These can be bosonized using:
\begin{equation}
    \hat\psi_{R/L}(x) = \frac{\hat\eta_{R/L}}{\sqrt{2\pi a}}\exp[\mp \ii\hat\phi(x) - \ii\hat\theta(x)],
\end{equation}
where $\hat\phi(x)$ and $\hat\theta(x)$ are dual bosonic fields satisfying $[\hat\phi(x),\partial_y\hat\theta(y)]=i\pi\delta(x-y)$, and $\hat \eta_{R/L}$ are Klein factors ensuring proper anticommutation relations.

The final Luttinger liquid Hamiltonian takes the form:
\begin{equation}
    \hat H = \frac{v}{2\pi}\int dx \left[K(\partial_x\hat\theta)^2 + \frac{1}{K}(\partial_x\hat\phi)^2\right],
\end{equation}
where $v$ is the velocity of excitations (which we set to $v=1$ in later discussions), and $K$ is the Luttinger parameter determined by Bethe ansatz for the XXZ model:
\begin{equation}
    K = \frac{\pi}{2\arccos\Delta} \,.
\end{equation}

If we treat $\partial_x\theta$ as momentum $p$ and $\phi$ as its conjugate coordinate $q$, using the relation between Lagrangian density $\mathcal{L}=i p\partial_\tau q-H$, we obtain the action density in phase space:
\begin{equation}
\mathcal{S}_{XXZ} = \frac{1}{2\pi} \int dx \, d\tau \left[ i (\partial_\tau \phi)(\partial_x \theta) + \frac{1}{K} (\partial_x \phi)^2 + K (\partial_x \theta)^2  \right].
\end{equation}
Integrating out either $\theta$ or $\phi$ yields the single-variable action:
\begin{align}
\mathcal{S}_{XXZ}(\phi) &= \frac{1}{2\pi K} \int dx \, d\tau \left[  (\partial_x \phi)^2 + (\partial_{\tau} \phi)^2  \right] \\
\mathcal{S}_{XXZ}(\theta) &= \frac{K}{2\pi} \int dx \, d\tau \left[  (\partial_x \theta)^2 + (\partial_{\tau} \theta)^2  \right]
\end{align}

This Luttinger liquid description captures the low-energy physics of the XXZ model, including its gapless excitations and power-law correlations. The Luttinger parameter $K$ determines the decay exponents of various correlation functions, with $K=1$ corresponding to the free-fermion point ($\Delta_c=0$) of the XXZ model. 

Using the relation between fermion and spin-1/2 operators, we can derive the standard bosonization dictionary for spin operators, which will be essential for analyzing decoherence effects:
\begin{gather}
    \hat{\sigma}^z_j \simeq \frac{2a}{\pi}\partial_x\hat{\phi} + c_1(-1)^j\cos(2\hat{\phi})\nonumber\\
    \hat{\sigma}^+_j \simeq e^{\ii\hat{\theta}}\left[c_2(-1)^j + c_3\cos(2\hat{\phi})\right]
\end{gather}
Before bosonizing the interaction, we need to clarify the meaning of the action and the final state after decoherence. This can be described using the following path integral formulation, where an additional imaginary time evolution arising from decoherence can be understood as a boundary interaction at $\tau=0$. The path integral equation takes the form:

\begin{gather} 
\bra{\phi_0}_R\tilde{\bra{{\phi}_0}}_Le^{\mu \sum_{i=1}^{L} \hat{\sigma}_{L;i}^\alpha \hat{\sigma}_{L;i+1}^\alpha \hat{\sigma}_{R;i}^\alpha \hat{\sigma}_{R;i+1}^\alpha} e^{-\beta(\hat{H}_L\otimes\hat{\mathbb{I}}_R+\hat{\mathbb{I}}_L\otimes\hat{H}^*_R)}\ket{\tilde\phi_{-\beta}}_L{\ket{{\phi}_{-\beta}}}_R \nonumber\\
=\int^{[\phi_\beta,\tilde{\phi_\beta}]}_{[\phi_0,\tilde{\phi_0}]}\mathcal{D}\phi\mathcal{D}\tilde{\phi}\mathcal{D}\theta\mathcal{D}\tilde{\theta} e^{-\mathcal{S}[\theta,\tilde{\theta},\phi,\tilde{\phi}]}
\end{gather}
as a result $|\rho\rangle\rangle$ is given by
\begin{gather}
    |\rho\rangle\rangle=\int^{[\phi,\tilde{\phi}]}\mathcal{D}\phi\mathcal{D}\tilde{\phi}\mathcal{D}\theta\mathcal{D}\tilde{\theta} e^{-\mathcal{S}[\theta,\tilde{\theta},\phi,\tilde{\phi}]}\ket{\phi}\ket{\tilde{\phi}}
\end{gather}
where the intergration is perform from $\tau=-\infty$ to $\tau=0$, and we consider the density matrix  and reduced density matrix of the double state, the posistion of the decoherence line is located at $\tau=-\eta$, where $\eta$ is a infinitely small number, which needs careful consideration when we manage to calculate entanglement entropy of the system, special attention needs to be paid to its relative position with respect to the branch cuts on the multi-sheeted Riemann surface.

\section{Details of strong coupling/weak coupling duality}
\label{appendix:C}
The Kondo problem can be mapped to a quantum mechanical scattering problem. Through mode expansion and integrating out the $\tau$ component of the system, we obtain a (0+1)d effective field theory. The action leads to an equation of motion:
$(\partial^2_{\tau}-p^2-m^2)\theta(p,\tau)=0$, where $\theta(p,\tau)$ is the Fourier transformation of $\theta(x,\tau)$ and $m$ is a small regulator. The solution takes the form:
\begin{align}
\theta(p,\tau)=&\theta(p,0)(\cosh{\omega_p\tau}-\coth{\omega_p\beta}\sinh{\omega_p\tau})\nonumber\\
&+\theta(p,\beta)\frac{\sinh{\omega_p\tau}}{\sinh{\omega_p\beta}}
\end{align}
where $\omega_p=\sqrt{p^2+m^2}$. Given the periodicity $\beta$ of $\tau$, we have $\theta(p,\beta)=\theta(p,0)$. In the limit $\beta\rightarrow\infty$ and $m\rightarrow0$, the effective theory of $XX$ decoherence near the critical point maps to the standard Kondo problem action:
\begin{gather}
S[\theta_{L/R}]=\frac{K}{\pi}\int \frac{dp}{2\pi} |p|\theta_{L/R}(p)\theta_{L/R}(-p)+\mu\int dx \cos(2\theta_{L/R})
\end{gather}

Similarly, the critical behavior near the critical point of ZZ decoherence can be described by:
\begin{gather}
S[\phi_{L/R}]=\frac{1}{\pi K}\int \frac{dp}{2\pi} \left(|p|+\frac{2\pi Kp^2}{\Lambda_0}\right)\phi_{L/R}(p)\phi_{L/R}(-p)+\mu'\int dx \cos(4\phi_{L/R})
\end{gather}

We now demonstrate that the effective action of $XX$ decoherence under strong decoherence is dual to an action with only weak decoherence.
\emph{Since the left-right subsystems are decoupled, we will omit the $L/R$ labels when no ambiguity arises.}
For large $\mu$, the free energy is minimized at $\theta=(k+\frac{1}{2})\pi$ (where $k\in\mathbb{Z}$). However, domain walls must exist between different configurations, leading to:
\begin{gather}
\theta(x)=\sum_i\int^x dx'\epsilon_if(x'-x_i)
\end{gather}
Here, $\epsilon_i=\pm1$ represents the domain wall number, and $f(x)$ describes tunneling between different classical minima, with the constraint $\int_{-\infty}^{\infty} dx f(x)=\pi$. The function decreases rapidly away from zero.

The action in terms of domain wall numbers becomes:
\begin{gather}
    S_{XX}=\frac{K}{\pi}\int \frac{dp}{2\pi}\frac{\pi^2}{|p|}\sum_{i,j}\epsilon_i\epsilon_je^{-p(x_i-x_j)}+nS_{\text{dw}}
\end{gather}
where $n$ is the total number of domain walls and $S_{\text{dw}}$ is the action of a single domain wall, assuming different domain walls don't interact.

Using the Hubbard–Stratonovich (HS) transformation, we obtain:
\begin{gather}
    S_{XX}=\frac{1}{4\pi K}\int dp \frac{|p|}{2\pi}\tilde\phi(p)\tilde\phi(-p)+\ii\sum_i\epsilon_i\tilde\phi(x_i)+nS_{\text{dw}}
\end{gather}
In the canonical formulation of the HS transformation, we have $[\epsilon_i,\tilde\phi(x_j)]=i\delta_{i,j}$, hence $[\partial_x\theta(x),\tilde\phi(y)]=i\pi\delta(x-y)$. Here, $\tilde\phi$ is the dual variable of $\theta$ in the Luttinger liquid theory. For simplicity, we will drop the tilde and denote it as $\phi$.

The partition function can then be written as:
\begin{gather}
    \mathcal{Z}_{XX}=\sum_n\sum_{\epsilon_i=\pm}\int\mathcal{D}\theta e^{-\frac{1}{4\pi K}\int dp \frac{|p|}{2\pi}\phi(p)\phi(-p)}\frac{e^{-nS_{\text{dw}}}}{n!}\prod_{i=1}^n\int dx e^{i\epsilon_i\phi(x_i)}
\end{gather}
Summing over $\epsilon_i$ and $n$, we arrive at:
\begin{gather}
    \mathcal{Z}_{XX}=\int\mathcal{D}\phi e^{-\frac{1}{4\pi K}\int dp \frac{|p|}{2\pi}\phi(p)\phi(-p)+\tilde\mu \int dx \cos\phi}
\end{gather}
where $\tilde\mu=e^{-2S_{\text{dw}}}$, a classical solution for domain wall is $\theta(x)=k\pi+\arctan(\sinh(2\mu^{1/2}x))$, and it was shown in ref \cite{PhysRevX.13.021026}
 that the leading term of $S_{\text{dw}}=2\mu^{1/2}$, so $\tilde\mu=\tilde\mu_0e^{-4\mu^{1/2}}$.

\section{Analysis of the Critical Line}

In this Appendix, we analyze the behavior of the R\'enyi-2 correlation function at the critical line ($\Delta_c=0$) under both weak and strong decoherence regimes. We also explain the relationship between the calculation of entanglement entropy and the impurity problem.

\subsection{R\'enyi-2 Correlation Function at Weak Decoherence Regime}
\label{appendix:D1}
We first examine the behavior of the system under weak decoherence using perturbation theory. In this regime, we can expand the correlation function to the subleading order when the decoherence strength is small. The correlation function can be expressed as follows:
\begin{gather}
    \langle \cos\theta(x_1)\cos \theta(x_2)\rangle_d = \langle \cos\theta(x_1)\cos \theta(x_2)\rangle + \mu\int dy \langle \cos\theta(x_1) \cos2\theta(y)\cos\theta(x_2)\rangle
\end{gather}
In this equation, $\langle\cdot\rangle_d$ denotes the expectation value under weak measurement, while $\langle\cdot\rangle$ represents the expectation value in the free theory. The first term corresponds to the correlation in the absence of decoherence, while the second term accounts for the corrections due to weak decoherence, represented by the coupling strength $\mu$.

For the free theory, the general n-point correlation function follows a power-law behavior, which is a hallmark of critical systems:
\begin{gather}
    \langle \prod_i e^{\ii p_i\theta(x_i)}\rangle = \prod_{i<j}(x_i-x_j)^{\frac{p_ip_j}{2K}}
\end{gather}
Here, this correlation function is non-zero only when the neutrality condition $\sum_i p_i=0$ is satisfied. If this condition is not met, the correlation function vanishes identically, indicating that contributions from different momenta cancel each other out.

Next, we explicitly evaluate the integral term in our expression:
\begin{align}
    &\int dy \langle \cos\theta(x_1) \cos2\theta(y)\cos\theta(x_2)\rangle \nonumber\\
    &= \frac{1}{8}\int dy \langle e^{\ii\theta(x_1)} e^{-2\ii\theta(y)}e^{\ii\theta(x_2)}\rangle + \text{h.c.} \nonumber\\
    &\sim C_1|x_2-x_1|^{-1/2}\ln\frac{|x_2-x_1|}{a}
\end{align}
In this expression, $a$ represents a short-distance cutoff, and $C_1$ is a constant that captures the strength of the correlation. 

Combining these results, we find that the Rényi-2 correlation function of $\sigma_x$ in the perturbative regime can be expressed as:
\begin{gather}
     C^{\text{II}}_{X}(x_2-x_1) = |\langle\cos{\theta(x_1)}\cos{\theta(x_2)}\rangle_d|^2 \sim \frac{1}{|x_2-x_1|}\left(1+2C_1\mu\ln\frac{|x_2-x_1|}{a}\right)
\end{gather}
This expression indicates that the correlation function exhibits a power-law decay with an additional logarithmic correction due to weak decoherence.

\subsection{R\'enyi-2 Correlation Function at Strong Decoherence Regime}
\label{appendix:D2}
For the strong decoherence case, where the decoherence strength $\mu$ is large (with $p$ approaching 1/2), we employ perturbation theory in the dual picture. In this regime, the correlation function can be expressed in terms of domain walls, which are essential in capturing the effects of strong decoherence:
\begin{gather}
    \langle e^{-\ii\theta(x_1)}e^{\ii\theta(x_2)}\rangle_w=\langle e^{\ii\int dx\partial_x\theta(x)\mathcal{T}_{x_1,x_2}(x)}\rangle_w
\end{gather}
Here, $\mathcal{T}_{x_1,x_2}(x)$ is a step function that equals 1 when $x_1<x<x_2$ and 0 otherwise. This formulation indicates that the correlation is influenced by the presence of domain walls between the points $x_1$ and $x_2$, which arise due to the decoherence process.

The full expression after expanding in terms of domain walls is given by:
\begin{gather}
    \langle e^{\ii\int dx\partial_x\theta(x)\mathcal{T}_{x_1,x_2}(x)}\rangle_w=\exp\left(-\sum_n\sum_{\epsilon_i}\left[\frac{1}{4\pi K}\int dp \frac{|p|}{2\pi}\tilde\phi(p)\tilde\phi(-p)+\ii\sum_i(\epsilon_i\phi(x_i)+\epsilon_i\pi\mathcal{T}_{x_1,x_2}(x_i))+nS_{\text{dw}}\right]\right)/\mathcal{Z}_{XX}
\end{gather}
This expression takes into account the contributions from both the fluctuations of the field $\phi$ and the effects of the domain walls, denoted by $S_{\text{dw}}$. The partition function $\mathcal{Z}_{XX}$ normalizes the correlation function, ensuring that the contributions from the entire system are properly accounted for.

We can then derive the correlation function as follows:
\begin{gather}
   \langle e^{-\ii\theta(x_1)}e^{\ii\theta(x_2)}\rangle_w =\int\mathcal{D}\theta e^{-\frac{1}{4\pi K}\int dp \frac{|p|}{2\pi}\phi(p)\phi(-p)+\tilde\mu \int dx \cos(\phi(x)+\pi\mathcal{T}_{x_1,x_2}(x))}/\int\mathcal{D}\theta e^{-\frac{1}{4\pi }\int dp \frac{|p|}{2\pi}\phi(p)\phi(-p)+\gamma \int dx \cos\phi(x)}
\end{gather}
In this equation, $\tilde{\mu}$ represents the strength of the coupling due to the presence of domain walls, while $\gamma$ relates to the coupling in the absence of domain walls.

For the partition function, we have:
\begin{gather}
    \int\mathcal{D}\theta e^{-\frac{1}{4\pi }\int dp \frac{|p|}{2\pi}\phi(p)\phi(-p)+\gamma \int dx \cos\phi(x)}=\mathcal{Z}_0\left(1+\frac{\tilde{\mu}^2}{2}\int dy_1dy_2\langle \cos\phi(y_1)\cos\phi(y_2)\rangle\right)
\end{gather}
This expression shows how the partition function is modified by the presence of the coupling $\tilde{\mu}$, indicating that the correlations between different points in space are affected by the interactions represented by $\cos\phi$.

At the critical line $K=1$, where the scaling dimension of the operator $e^{\ii\phi(x)}$ is unity, we obtain:
\begin{align}
    \langle e^{-\ii\theta(x_1)}e^{\ii\theta(x_2)}\rangle_w &= 
    \frac{1+\frac{\tilde{\mu}}{4}\int d y_1dy_2\frac{1}{|y_1-y_2|^{2}}e^{\ii(\pi(\mathcal{T}_{x_1,x_2}(y_1)-\mathcal{T}_{x_1,x_2}(y_2)))}}{1+\frac{\tilde{\mu}}{4}\int d y_1dy_2\frac{1}{|y_1-y_2|^{2}}} \nonumber\\
    &\sim 1-C_2\tilde{\mu}^2\ln\frac{|x_2-x_1|}{a}
\end{align}
In this expression, $C_2$ is a constant that quantifies the strength of the logarithmic correction. The result indicates that the correlation function approaches 1 in the limit of large separation between $x_1$ and $x_2$, but exhibits a logarithmic decay due to strong decoherence.

Finally, the Rényi-2 correlation function of $\sigma_x$ in the strong decoherence regime is given by:
\begin{gather}
     C^{\text{II}}_{X}(x_2-x_1) = |\langle\cos{\theta(x_1)}\cos{\theta(x_2)}\rangle_d|^2 \sim 1-2C_2\tilde{\mu}^2\ln\frac{|x_2-x_1|}{a}
\end{gather}
This result demonstrates that strong decoherence leads to a logarithmic decay of correlations, contrasting with the power-law behavior modified by logarithmic corrections observed in the weak decoherence regime. This distinction highlights the significant impact of decoherence on the correlations within the system, emphasizing the transition from coherent to incoherent behavior as the decoherence strength increases. This also provides another perspective on why $\Delta_c=0$ is a critical point: when decoherence is strong, the order parameter experiences a logarithmic decrease relative to 1. If the decrease is faster, the order becomes unstable; if the decrease is slower than logarithmic, it indicates that there is order.

\subsection{Map of decoherence problem to impurity problem}
\begin{figure*}[htbp]
\centering
\includegraphics[width=0.32\textwidth]{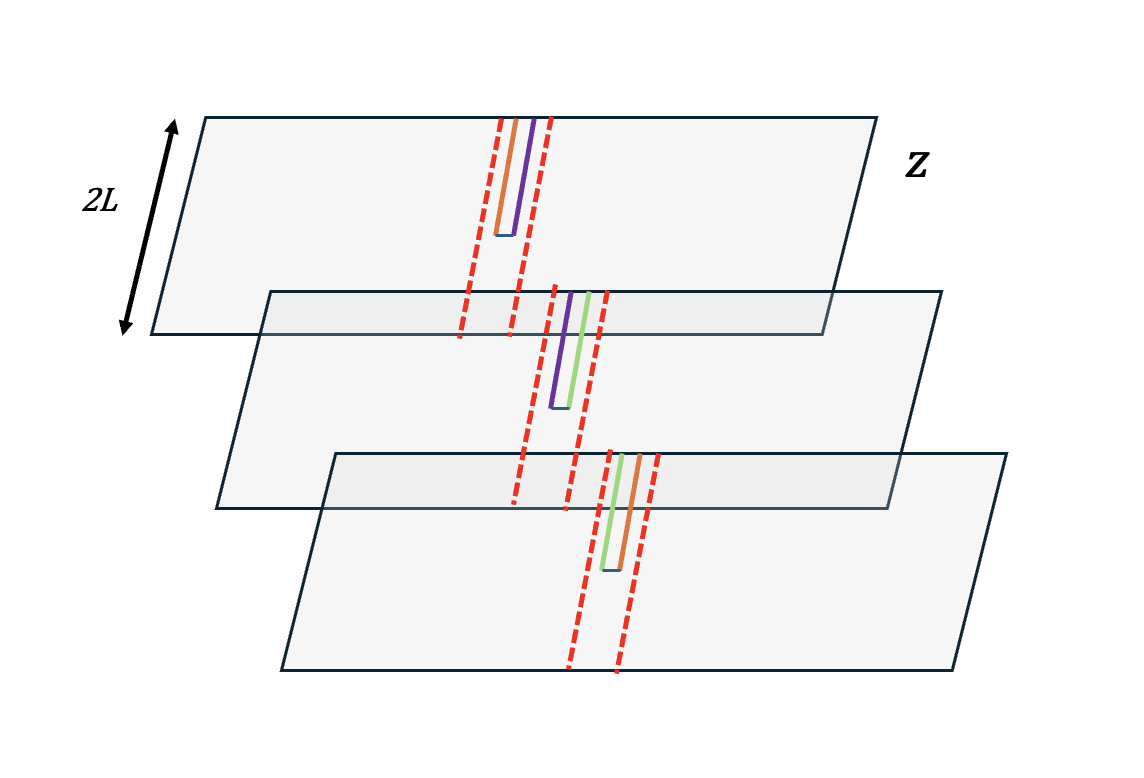}
\hfill
\includegraphics[width=0.32\textwidth]{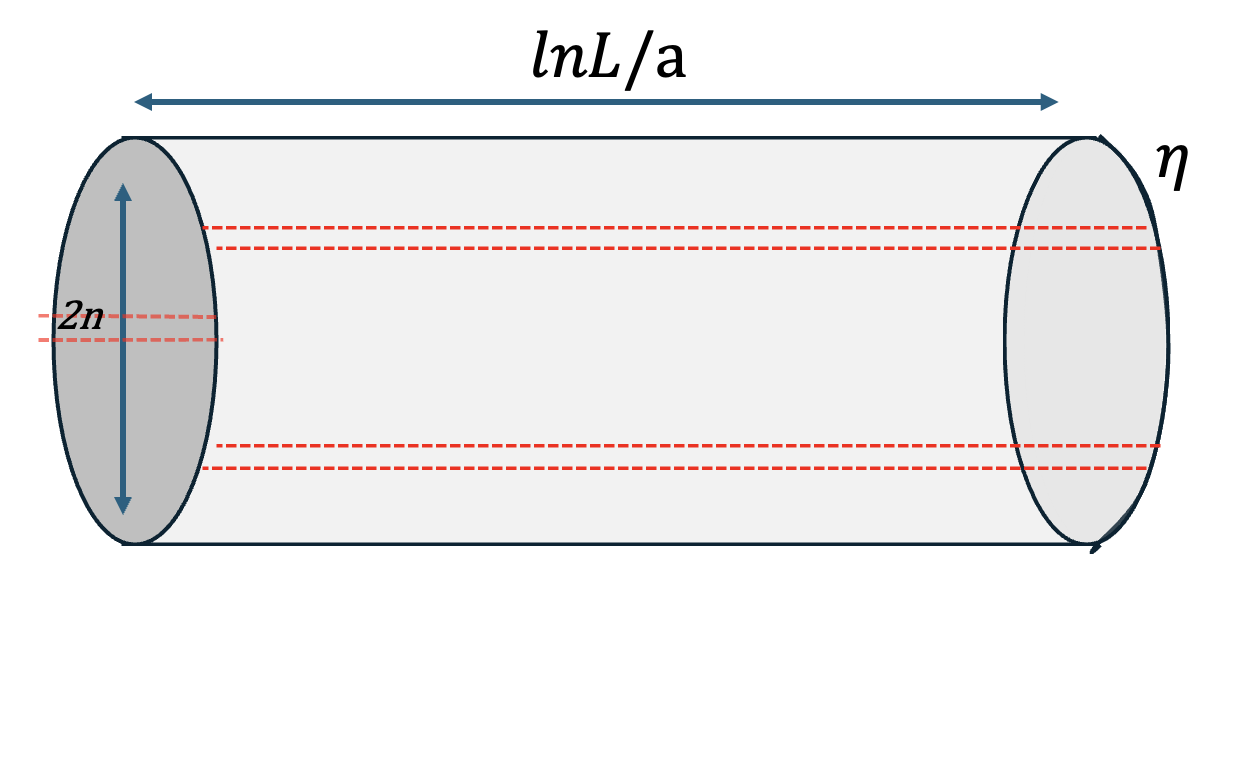}
\hfill
\includegraphics[width=0.32\textwidth]{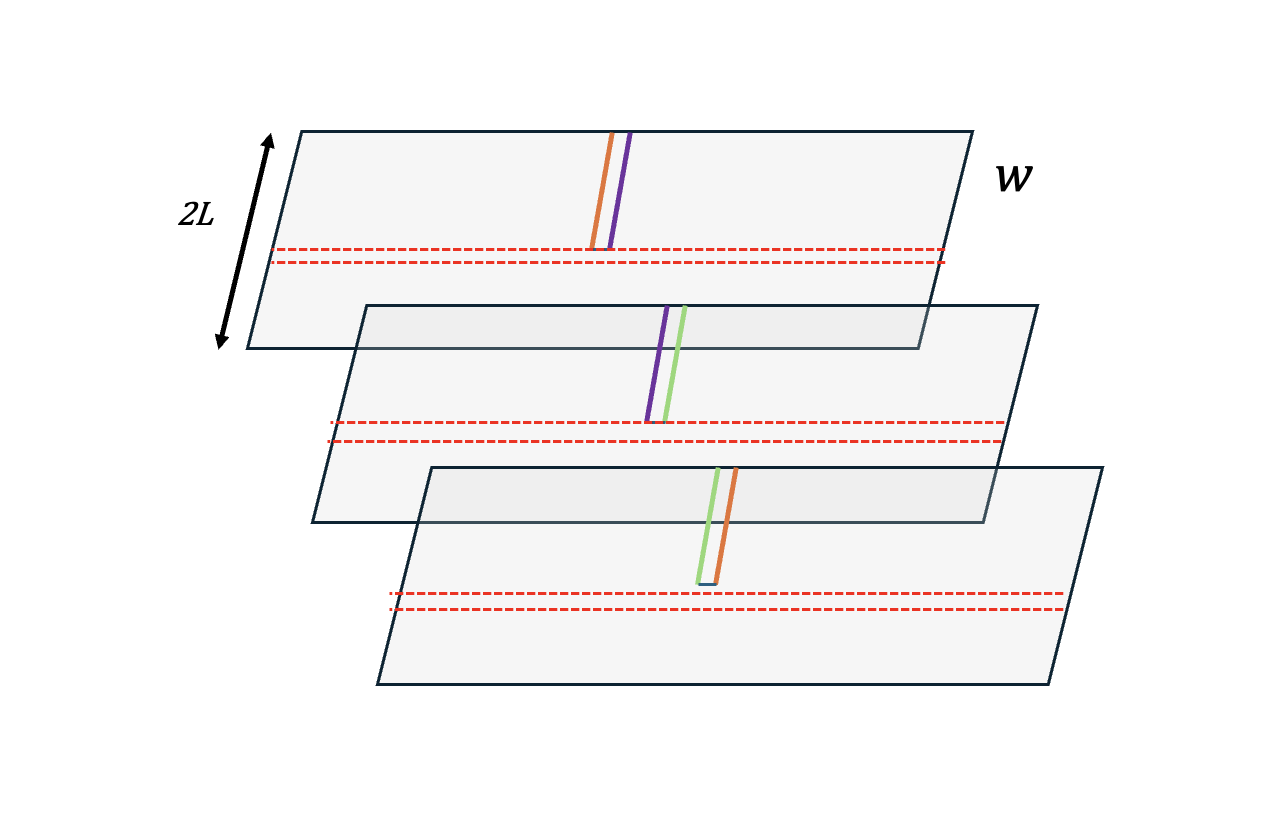}
\caption{An illustration of our conformal transformation for $n=3$. The colorful solid lines represent branch cuts, with branch cuts of the same color connected to each other. The dashed lines indicate defect lines.}
\label{conformal}
\end{figure*}
\label{appendix:D3}
 As discussed in the main text, the half-chain entanglement entropy of the left/right piece of the Choi state can be mapped to the left-right entanglement entropy of a Dirac fermion with a mass defect at the center of the chain. Here, we detail the conformal map employed in this analysis.To calculate the R\'enyi-$n$ entropy, we perform a path integral on an $n$-sheet Riemann surface. Each sheet of this surface is represented by a finite-length stripe where $-L<\text{Im}~z<L$ and $-\infty<\text{Re}~z<\infty$. A branch cut exists at $\text{Im}~z>0$, while the decoherence effect manifests as a defect line localized at $\text{Re}~z=\pm 0^+$.We first introduce the transformation $\eta=\ln z$, which maps our system to a Riemann surface with the periodicity condition $\text{Im}~\eta\sim\text{Im}~\eta+2n\pi$, and when $L$ is large enough ( $L\gg a$ ) \cite{Brehm2015}, we assume that path integral on a stripe with width $2L$ won't have large distinguish with disk $|z|<L$,  it can be regarded as a cylinder with height $\ln\frac{L}{a}$, where $a$ is the cut-off of lattice . In this representation, the defect lines are localized at $\text{Im}~\eta=(k+\frac{1}{2})\pi$ for $k=0,1,\ldots,n-1$. Subsequently, we apply a translation followed by an exponential mapping: $w=\exp(\eta-\ii\pi/2)$. This transformation yields an $n$-sheet Riemann surface where the defect line is positioned along $\text{Im}~w=0$. Then following the calculation of free fermion in Ref \cite{eisler2010entanglement}, we can reach the result 
\begin{align}
    c_{\text{eff}} = -\frac{6}{\pi^2} \bigg\{ [(1+s) \log(1+s) + (1-s) \log(1-s)] \log(s) 
    + (1+s) \text{Li}_2(-s) + (1-s) \text{Li}_2(s) \bigg\}.
\end{align}
in the main text, we have shown that the entanglement entropy and the system size exhibit a good numerical logarithmic relationship, which also indicates that our approximation is reasonable.

\section{Details of the matrix product state simulation}
\label{appexdix:E}
In this appendix, we provide some details of the numerical simulations. 
Firstly, the ground state $|\Psi_{0}\rangle$ of the spin-$1/2$ XXZ model is solved by the conventional finite-size density matrix renormalization group (DMRG) algorithm~\cite{white1992prl,white1993prb} using matrix product state (MPS) representations~\cite{cirac2006prb}. 
\begin{figure}[htbp]
    \centering
    \includegraphics[width=0.75\linewidth]{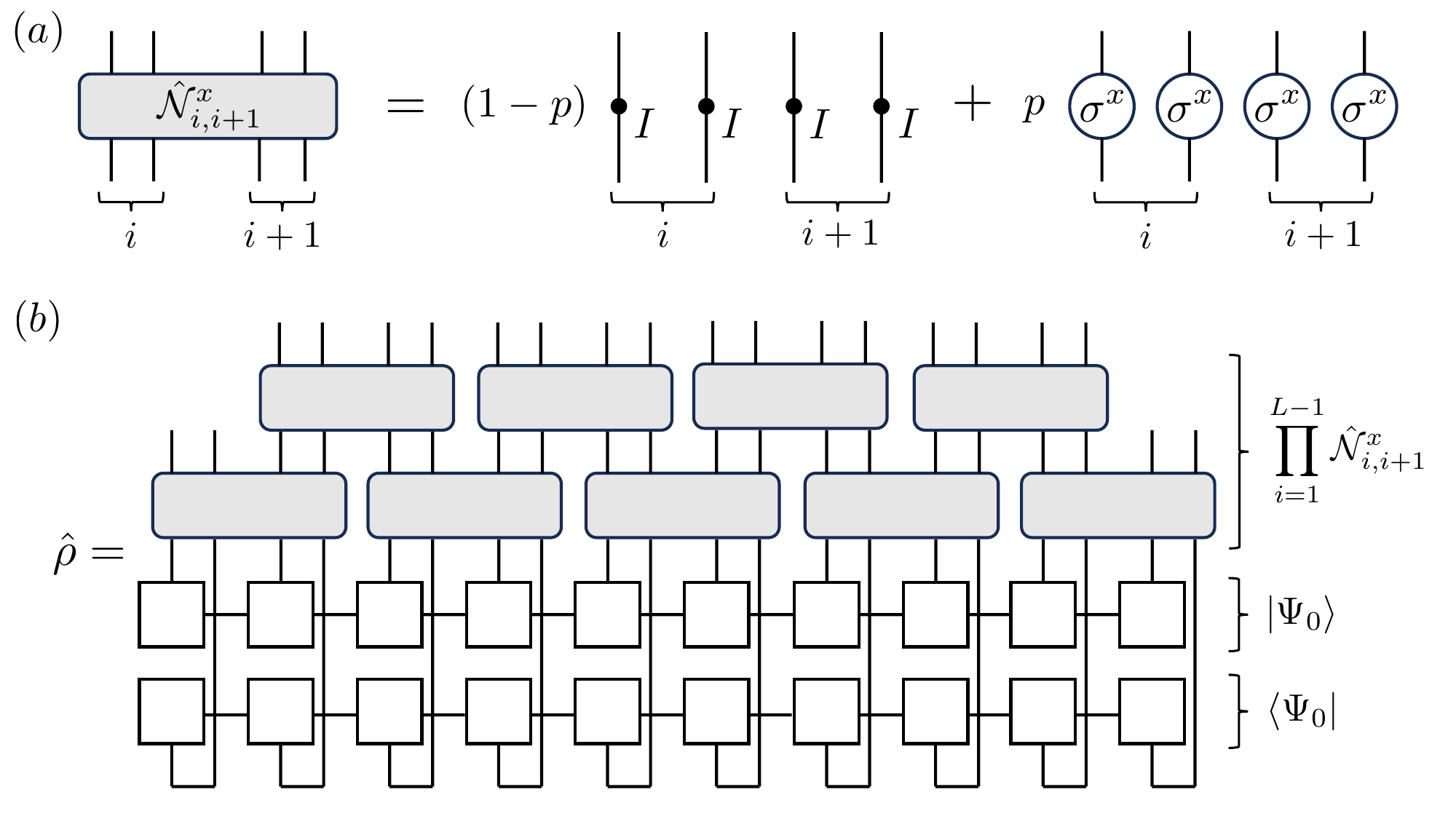}
    \caption{(a) The construction of the local decoherence channel $\hat{\mathcal{N}}_{i,i+1}^{x}$. The channel for the ZZ decoherence can be obtained by replacing $x$ with $z$.
    (b) The tensor-network diagram of the decohered mixed state $\hat{\rho}$. 
    }
    \label{fig:sm_channel}
\end{figure} 
The accuracy of the algorithm is generally controlled by the MPS bond dimension $\chi$. 
Then we can apply the local decoherence channels to the density matrix of the ground state, i.e., $\hat{\rho}_{0} = |\Psi_{0}\rangle\langle{\Psi_{0}}|$, to obtain the decohered mixed state $\hat{\rho}$. 
Concretely, the local decoherence channel $\hat{\mathcal{N}}_{i,i+1}^{\alpha}$ can be written as a rank-$8$ tensor as shown in Fig.~\ref{fig:sm_channel} (a) and the final decohered mixed state is represented by the tensor network diagram in Fig.~\ref{fig:sm_channel} (b). 
However, to evaluate the quantities concerned in our work, we do not directly calculate the density matrix $\hat{\rho}$ or save it in the hard disk. 
Instead, the various R\'enyi-2 type quantities, such as the R\'enyi-2 correlator and the R\'enyi-2 entropy, are computed by contracting two copies of the network diagram shown in Fig.~\ref{fig:sm_channel} (b) with properly inserted local operators~\cite{zou2023prl}. 
Due to the commutativity of the local channels, the contraction can be performed recursively from left to right with time complexity $O(L\chi^{5})$. 
In this work, the MPS bond dimension is set to $\chi=100$ unless otherwise specified. 
We would like to note that the contraction process for the calculation of the quantities is strictly exact, therefore, the only limitation to the precision of the results is the finite $\chi$. 

\begin{figure}
    \centering
    \includegraphics[width=0.75\linewidth]{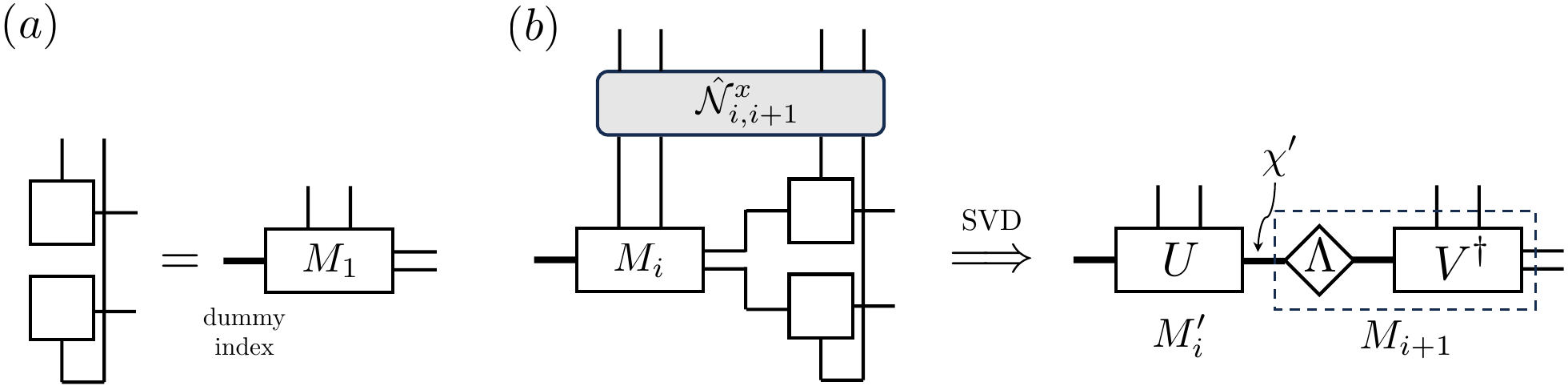}
    \caption{Bring the doubled state $|\rho\rangle\rangle$ into canonical form in an iterative way. We assume that the orthogonality center of $|\Psi_{0}\rangle$ has been brought to the first site. The final compressed doubled state is represented by a series of rank-$4$ tensors $M'_{i}$. (a) The construction of the first tensor $M_{1}$ where the thick leg is a dummy index of dimension $1$. (b) The recursive relation to obtain the next tensor $M_{i+1}$ from $M_{i}$ and $\hat{\mathcal{N}}_{i,i+1}^{x}$. The SVD guarantees that the obtained $M'_{i}$ is left-normalized. The thick link between tensors denotes a truncation bond introduced in the SVD whose dimension is $\chi'$.}
    \label{fig:sm_svn}
\end{figure}

In our work, we also investigate the half-chain von Neumann entanglement entropy of the doubled state $|\rho\rangle\rangle$. 
For this purpose, we choose to bring the state $|\rho\rangle\rangle$ in canonical form as explained in Fig.~\ref{fig:sm_svn}. 
The resulting compressed state is represented by a series of rank-$4$ tensors $M'_{i}$ and the von Neumann entropy can be computed using the diagonal matrices $\Lambda_{i}$ after proper renormalization. 
Here, another bond dimension $\chi'$ is introduced in the singular value decomposition (SVD), which is denoted by the thick link between local tensors. 
In practice, we set $\chi' = m\chi$ and find $m \ge 6$ is large enough to obtain reliable results in our case.

Finally, the simulation of the weak measurement on the gound state, $\hat{K}_{\alpha} |\Psi_{0}\rangle$, is quite straightforward and we refer interested readers to Ref.~\cite{ulrich2011aop}. 
It is noted that, in this case, the value of the MPS bond dimension is not limited by the high computational complexity (such as $O(L\chi^{5})$ in the calculation of R\'enyi-2 correlators), therefore, we can employ large enough $\chi$ to achieve accurate results in the simulation of $\hat{K}_{\alpha} |\Psi_{0}\rangle$ for $L$ up to $512$.

\section{Other Numerical Results}
\label{appendix:F}
\begin{figure*}[htbp]
     \centering
    \includegraphics[width=0.75\linewidth]{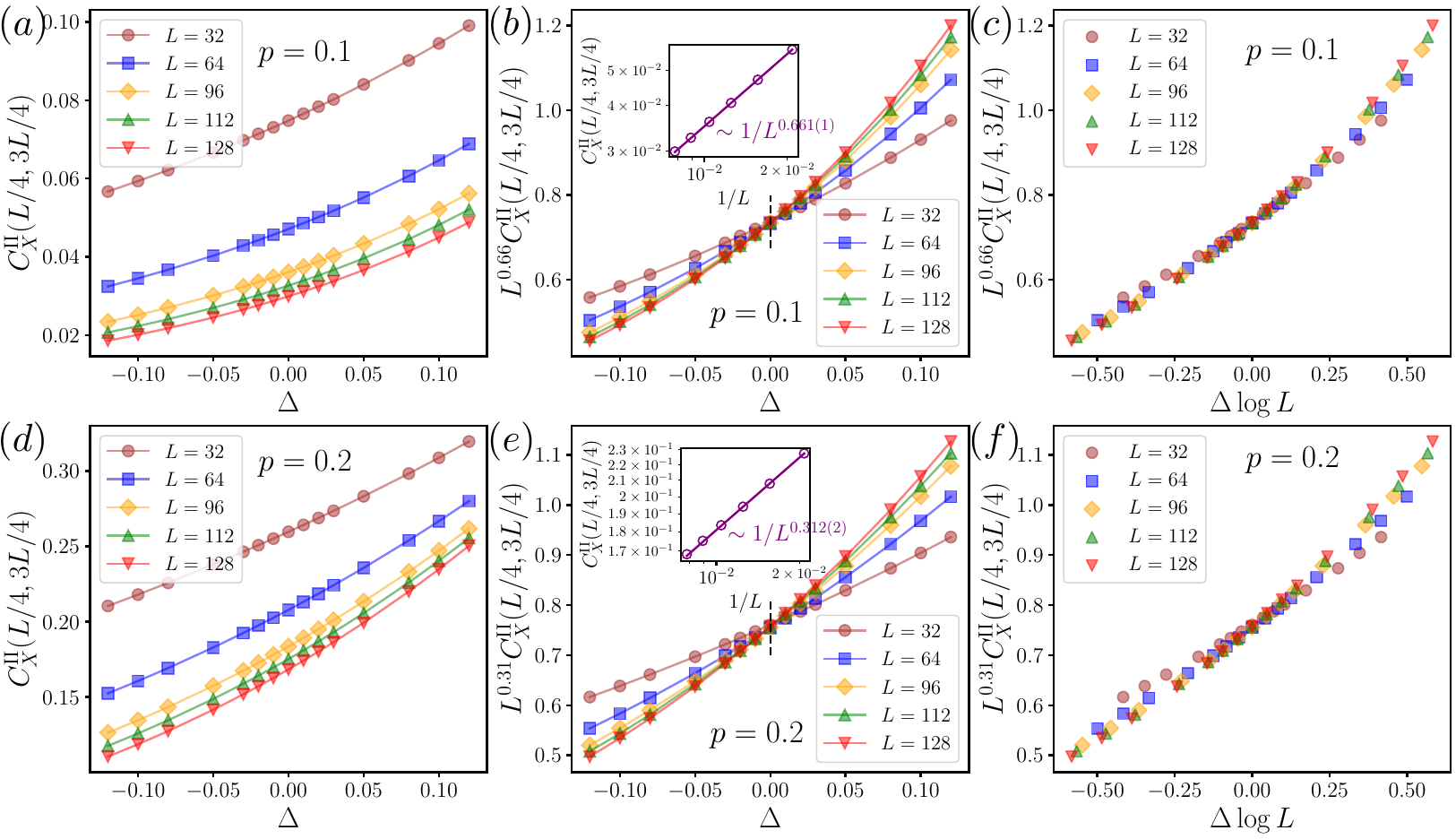}
    \caption{Data collapse for (a)-(c) \( p=0.1 \) and (d)-(f) \( p=0.2 \). The finite-size scaling behavior shows excellent agreement with the expectation from the boundary BKT transition. The least-squares fitting shown in the insets of (b) and (e) gives \( \eta=0.661 \) for \( p=0.1 \) and \( \eta=0.312 \) for \( p=0.2 \)\,. Simulated system size is $L=32$ to $128$ here.}
    \label{collaps}
\end{figure*}

In the main text, we claim that for different values of \( p \), there exists a boundary BKT phase transition at \( \Delta_c = 0 \). Furthermore, the critical exponent associated with the phase transition decreases as \( p \) increases. Here, we present the numerical results for \( p = 0.1 \) and \( p = 0.2 \). The finite-size scaling behavior shows excellent agreement with the data collapse described by \( \Delta \log L \). Additionally, the critical exponent $\eta$ in \( C^{\text{II}}_{X}(L/2) \sim L^{-\eta} \) also exhibits a numerical decrease as seen in the insets of Figs.~\ref{collaps} (b) and (e).

\twocolumngrid
\let\oldaddcontentsline\addcontentsline
\renewcommand{\addcontentsline}[3]{}
\bibliography{main.bib}
\let\addcontentsline\oldaddcontentsline

\end{document}